\documentclass[twocolumn,showpacs,amsmath,amstex,amssymb,mathfonts,prb]{revtex4-1}
\usepackage{graphicx,bm,units}
\usepackage{enumerate}
\usepackage{multirow}
\usepackage{color}

\newcommand{\C}{${\bf C}\textrm{ }$}

\def\ket#1{{|  #1 \rangle}}

\newcommand{\bs}{\backslash}

\begin{document}

\title{Bulk-Edge Correspondence in the Entanglement Spectra}

\author{A. Chandran$^1$, M. Hermanns$^1$, N. Regnault$^2$, and B.A. Bernevig$^1$}
\affiliation{$^1$ Department of Physics, Princeton University, Princeton, NJ 08544} 
\affiliation{$^2$ Laboratoire Pierre Aigrain, ENS and CNRS, 24 rue Lhomond, 75005 Paris, France}

\begin{abstract}  Li and Haldane conjectured and numerically substantiated that the entanglement spectrum of the reduced density matrix of \emph{ground-states} of  time-reversal breaking  topological  phases (fractional quantum Hall states) contains information about the counting of  their edge modes when the ground-state is cut in two spatially distinct regions and one of the regions is traced out.  We analytically substantiate this conjecture for a series of FQH states defined as unique zero modes of pseudopotential Hamiltonians by finding a one to one map between the thermodynamic limit counting of two different entanglement spectra: the particle entanglement spectrum, whose counting of eigenvalues for each good quantum number is identical  (up to accidental degeneracies) to the counting of bulk quasiholes, and the orbital entanglement spectrum (the Li-Haldane spectrum). As the particle entanglement spectrum is related to bulk quasihole physics and the orbital entanglement spectrum is related to edge physics, our map can be thought of as a mathematically sound microscopic description of bulk-edge correspondence \emph{in entanglement spectra}. By using a set of clustering operators which have their origin in conformal field theory (CFT) operator expansions, we show that the counting of the orbital entanglement spectrum eigenvalues in the thermodynamic limit must be identical to the counting of quasiholes  {\emph {in the bulk}}. The latter equals the counting of edge modes at a hard-wall boundary placed on the sample. Moreover, we show this to be true even for CFT states which are likely bulk gapless, such as the Gaffnian wavefunction.  
\end{abstract}


\date{\today}

\maketitle
\section{Introduction}

Determining the universality class of a real system exhibiting a topological phase is a difficult task in condensed matter physics. 
Renormalization group methods have been very successful in uncovering the universal physics in phases with local order parameters, but, due to their perturbative approach, cannot be readily generalized to topological phases  which do not exhibit symmetry breaking.  The density matrix renormalization group\cite{PhysRevLett.69.2863,RevModPhys.77.259, shibata2001prl5755, PhysRevLett.100.166803} and tensor matrix product states\cite{Vidal07,Verstraete04} can probe topological order in one dimension, but have had limited success with higher dimensional systems so far. 
The prototype of two-dimensional topologically ordered phases are the experimentally accessible fractional quantum Hall (FQH) phases. A promising tool to extract topological information from the ground state wavefunction in these phases is the entanglement entropy\cite{levinwen, kitaevpreskill,calabrese-04jsmp06002,pasquier,Dong-jh}. However, it depends on scaling arguments, is hard to obtain to sufficient accuracy from numerical calculations\cite{haque2007,lauchli-10njp075004}, and does not uniquely determine the topological order in the state.

In 2008, Li and Haldane\cite{Li:2008aa} proposed a new tool to identify topological order in non-abelian FQH states -- the entanglement spectrum. They divided the single-particle orbitals in a Landau level on the sphere along the equator and constructed the reduced density matrix of the ideal (model) and the realistic (Coulomb) FQH states in the upper half of the sphere (part $A$) by tracing out orbitals in the lower half (part $B$). Having thus created a `virtual' edge, they defined the orbital entanglement spectrum (OES) to be the plot of the negative logarithm of the eigenvalues  of the reduced density matrix of $A$ vs the $z$-angular momentum of $A$ ($L_z^A$) for a fixed number of particles in $A$.   In particular, Li and Haldane considered the part of the spectrum with the lowest-lying levels and the highest-weight eigenstates of the reduced density matrix of $A$. 
They noticed that the number of levels in every OES of the model states, such as the Laughlin and the Moore-Read, was much smaller than the Hilbert space dimension and was identical to the counting of the conformal field theory (CFT) modes  associated with the edge at small values of $L_z^A$. Although the number of levels in the OES of the Coulomb state saturated the Hilbert space dimension, a gap separated the levels higher in the spectrum from a CFT-like low-lying spectrum at small values of $L_z^A$ with the \emph{same} counting as the model state. This was taken as evidence that the Coulomb state at $\nu=5/2$ and the model Pfaffian state belonged to the same universality class. Based on extensive numerical evidence, they conjectured that --- 1) In the thermodynamic limit, the counting of the OES of the model state is the counting of the modes of the conformal theory describing its gapless edge excitations, 2) The `entanglement gap' separating the low-lying, CFT-like levels
  from the generic ones higher in the Coulomb  spectrum is finite in the thermodynamic limit.

\begin{figure*}[htbp]
     \includegraphics[width=0.75 \textwidth]{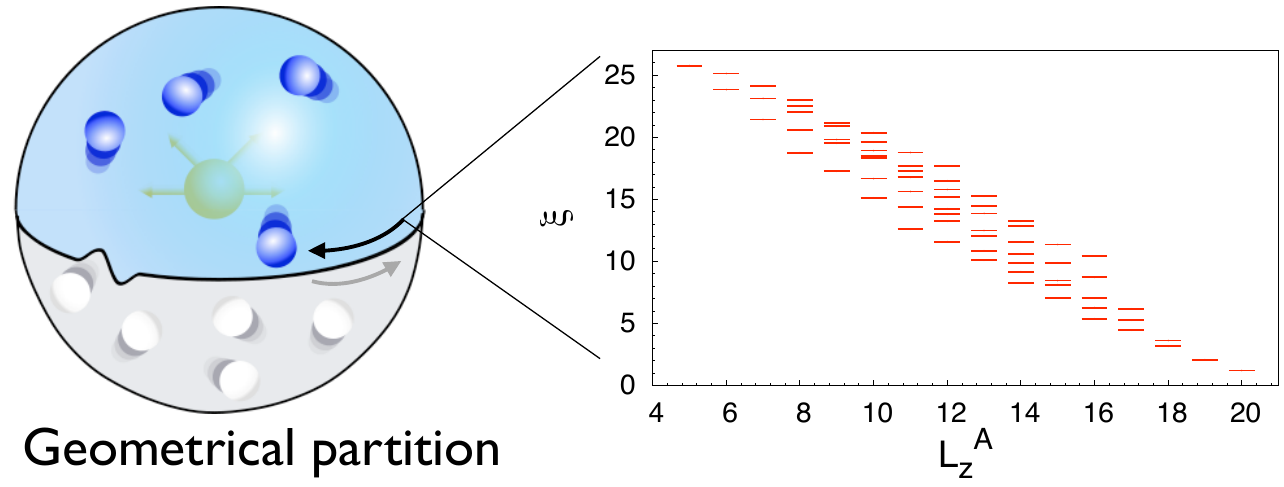}
         \caption{
Left: Sketch of the partition in orbital space (part $B$ in grey). The tracing procedure creates a virtual edge and the orbital entanglement spectrum (OES) probes the chiral edge mode(s) of part $A$. Right: OES of the $\nu=1/2$ Laughlin state of $N=9$ bosons, with orbital cut $l_A=8$ and $N_A=4$. The minimal angular momentum $L^A_{z,min}$ defined in the text is the $L^A_{z,min}=20$ sector in the plot. The entanglement level counting at $L_z^A=20,19,\ldots,16 $ is $(1,1,2,3,5 )$ which is the counting of modes of a $U(1)$ boson in the thermodynamic limit. Finite size effects appear at $L_z^A=15$. }
      \label{fig:oem_n_9_2s_16_na_4_la_8}
\end{figure*}

Many researchers have investigated properties of the entanglement spectra since. The authors of [\onlinecite{thomale2010}] discovered that the entanglement spectrum in the thin-annulus limit (the conformal limit) had  for, several examples,  a \emph{full} gap at finite system sizes. The counting of the \emph{entire} low-lying spectrum of the Coulomb state is the same as that of the corresponding model state in this limit. Motivated by this result, we recently conjectured a counting principle for the finite-size counting of the OES of the Laughlin states\cite{hermanns2010}. Other cuts have also been studied. Tracing out a fraction of the particles in the many-body ground state corresponds to a particle cut; the entanglement spectrum of the resulting reduced density matrix is the particle entanglement spectrum (PES) introduced in [\onlinecite{sterdyniak:2010aa}]. The level counting of the PES of a model state (described as 
 a CFT correlator) is bounded from above by the number of \emph{bulk} quasi-hole states of the model state; along with the OES, it is conjectured to contain all the topological numbers of the state. Entanglement spectra in other systems have also been explored; see, for instance, Refs.~[\onlinecite{Hughes2010,  fidkowski2011, Hughes2010Inv, turner2010, PhysRevLett.105.080501, Fagotti2010, Franchini2010, turnerarx1008, pollmann2010njp025006, pollmann2010prb064439, didier2010prl077202, thomale2010prl116805, calabrese2008pra032329, lauchli10,  zozulya2009prb045409, papicarx1008, Liu2010, schliemannarx1008}].

\begin{figure*}[htbp]
     \includegraphics[width=0.75 \textwidth]{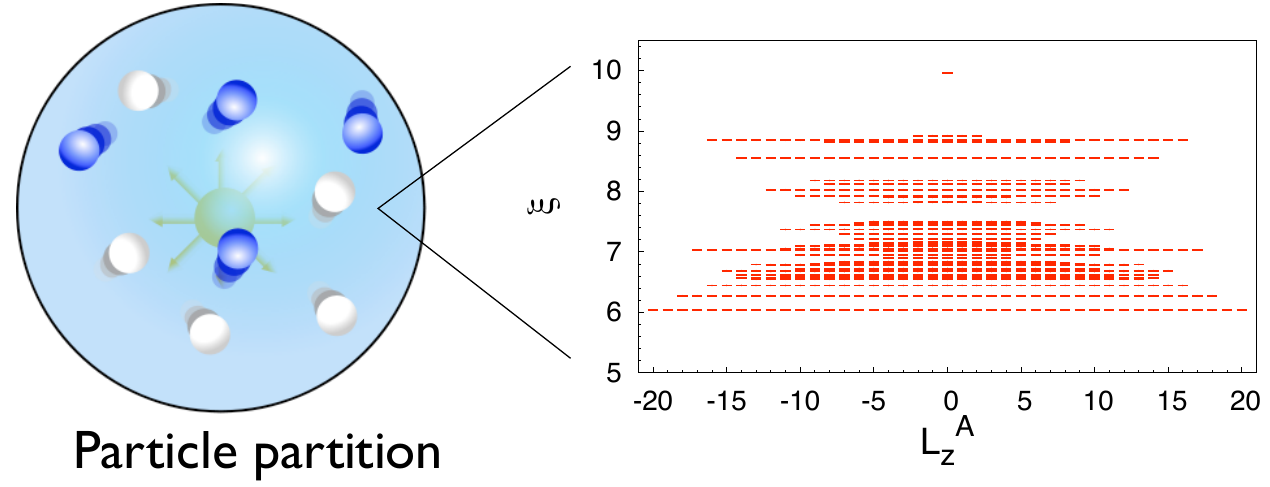}
         \caption{
    Left: Sketch of the partition in particle space. The particles of part $B$ that are traced out are denoted in grey. Right: particle entanglement spectrum (PES) of the $\nu=1/2$ Laughlin state of $N=9$ bosons, with particle cut $N_A=4$. The minimum angular momentum $L^A_{z,min}$ defined in the text is $L^A_{z,min}=20$  in the plot.
      The entanglement level counting at $L_z^A=20,19, \ldots,15 $ is $(1,1,2,3,5,6) $, identical to the counting of quasiholes in a Laughlin state of $N_A$ particles but at flux $N_\phi = 16$.
         }
      \label{fig:pem_n_9_2s_16_na_4_la_8}
\end{figure*}

Analytic work in this emerging field is challenging because of the strongly interacting nature of  FQH states. The Li-Haldane conjecture (the correspondence between the counting of the number of modes in the thermodynamic limit OES and the counting of the edge-excitation spectrum) is easy to prove in non-interacting systems, such as the Integer Quantum Hall system and topological insulators\cite{PhysRevLett.104.130502,turner2010prb241102,peschel}. In this article, we partially prove the first part of the Li-Haldane conjecture for clustering model states: in the thermodynamic limit, we show that the counting of the low-lying levels of the OES is identical to the PES counting and is bounded above by the counting of the CFT associated with the edge. We prove this bound for the bosonic $(k,2)$-clustering states (the Read-Rezayi sequence) multiplied by any number of Jastrow factors, and for the Gaffnian. In principle, the bound should hold for all model states defined as the unique, highest-density zero modes of $(k+1)$-body pseudopotential Hamiltonians\cite{Simon:2007kx}. The starting point of the proof is at the PES. We show that the counting of the PES is bounded above by the number of zero modes of the model pseudopotential Hamiltonian (number of bulk q-h excitations) of the particles in $A$\cite{sterdyniak:2010aa}. We then establish a bulk-edge correspondence in the entanglement spectrum by showing that the particle and orbital entanglement spectra have the same counting. This correspondence holds for a part of every spectrum at finite size and for the entire spectrum in specific quantum number sectors in the thermodynamic limit, as will be made precise later. Using the known bulk-edge correspondence in the energy spectrum of bulk gapped FQH states, we analytically prove that the counting of the natural OES is bounded above by the counting of the CFT describing the edge, and reason physically that the bound should be saturated. 

The paper is organized as follows: We define the orbital and particle entanglement matrices and spectra in Sections~\ref{Sec:OEM} and \ref{sec:pemintro}. In Sec.~\ref{Sec:Ptilde}, we present the upper bound to the number of levels in the particle entanglement spectrum and argue for its saturation. In Sec.~\ref{Sec:Squeezingconstraints}, we formulate the clustering properties of the model state in the single-particle orbital basis and use them to relate the counting of the particle and orbital entanglement spectra of the Read-Rezayi sequence in Sec.~\ref{Sec:Theproof}. The proof for the upper bound of the Li-Haldane conjecture is presented at the end of the same section. In Sec.~\ref{Sec:Otherstates}, we extend the proof to the other model states. Examples and the mathematical formulation of the ideas in the proof are in the Appendices.

\begin{figure*}[htbp]
     \includegraphics[width=0.75 \textwidth]{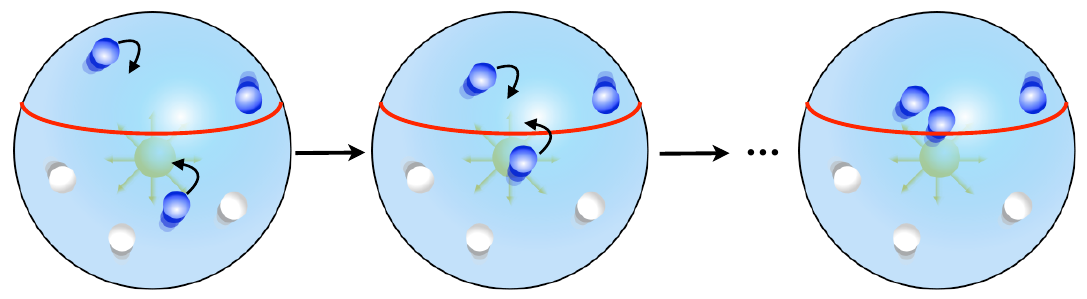}
         \caption{
The configurations in the PES can be related to those of the OES using the clustering constraints. These constraints reveal the vanishing properties of the FQH state as particles are brought closer together. They relate the long-wavelength properties of the FQH state when two particles are far away from each other to the short-wavelength  properties of the state when particles are close together, and hence can be used to 'drag' particles from the PES hilbert space into the more restrictive OES hilbert space. 
         }
      \label{fig:pemtooem}
\end{figure*}

\section{Notation}
\label{Sec:Notation}
The results that we present in this article hold on any surface of genus 0 (such as the disc or the sphere) pierced by $N_\phi$ flux quanta; for simplicity, we choose the sphere geometry. The single-particle  states of each Landau level are eigenstates of $\hat{L}_z$, the $z$-component of angular momentum and $|\vec{L}|^2$, the square of the magnitude of the total angular momentum vector\cite{Haldane:1983aa}. In the Lowest Landau Level (LLL), the degenerate single-particle  states belong to a multiplet of angular momentum $L=N_\phi/2$ and consequently, $L_z\in [-N_\phi/2, \ldots, N_\phi/2]$. We are forced to adopt a dual notation in this article - the single-particle  orbitals are indexed by their $z$-angular momentum $L_z$ in the figures, and by a shifted label $(N_\phi/2-L_z)$ in the text. At the north (south) pole, $L_z=N_\phi/2$ ($-N_\phi/2$).

Fermionic/bosonic many-body wavefunctions of $N$ particles and total angular momentum $L_z^{tot}$ can be expressed as linear combinations of Fock states in the occupation number basis of the single-particle  orbitals. Each Fock state $|\lambda\rangle$ can be labeled either by $\lambda$ or by $n(\lambda)$. $\lambda =[\lambda_1, \lambda_2, \ldots, \lambda_{N}]$ is an ordered partition of $L_z^{tot}$ into $N$ parts; each orbital with index $\lambda_j$ is occupied in the Fock state.  By definition, $\lambda_i\geq\lambda_j \textrm{ if } i<j$. $n(\lambda)$ is the occupation number configuration. It is defined as $n(\lambda)=\{n_j(\lambda), j=0,\ldots N_\phi\}$, where $n_j(\lambda)$ is the occupation number of the single-particle  orbital with angular momentum $j$. For instance, if $N_\phi=6$, the partition and occupation configuration of a Fock state $|4,2,0\rangle$ of $3$ particles is $\lambda=[4,2,0]$ and $n(\lambda)= \{1010100\}$ respectively. We will repeatedly run into a special kind of partition in this article --- the $(k,r)$-admissible partition.  A $(k,r)$-admissible partition labels a Fock state, whose occupation configuration has no more than $k$ particles in $r$ consecutive orbitals. The partition in the example above is $(1,2)$-admissible.

Three useful relations between partitions are `dominance', `squeezing' and `addition'. 
A set of partitions may always be partially ordered by dominance, indicated by the symbol `$>$'. A partition $\mu$ dominates another partition $\nu$  ($\mu>\nu$) iff $\sum_{i=0}^{r} \mu_i \geq \sum_{i=0}^{r} \nu_i \,\forall r \in[0, \ldots, N]$. 
Squeezing is a two-particle operation that connects $n(\mu)$ to $n(\nu)$. It modifies the orbitals occupied by any two particles in $n(\mu)$ from $m_1$ and $m_2$ to $m_1'$ and $m_2'$ in $n(\nu)$, such that $m_1+m_2=m_1'+m_2'$ and $m_1<m_1'\leq m_2'<m_2$ if the particles are bosonic or $m_1<m_1' < m_2'<m_2$ if they are fermionic. Dominance and squeezing are identical concepts: a partition $\mu$ dominates a partition $\nu$ iff $\nu$ can be squeezed from $\mu$ by a series of squeezing operations. 
The `sum' of two partitions $\mu+\nu$ is defined as the partition with occupation configuration $n(\mu+\nu)=\{n_j(\mu)+n_j(\nu), j=0,\ldots N_\phi\}$.

FQH wavefunctions in the LLL are translationally invariant, symmetric, homogeneous polynomials of the coordinates of the $N$ particles, ($z_1,z_2\ldots z_N$). 
We consider mainly the bosonic Read-Rezayi sequence at filling $\nu=k/2$ here (see Sec.~\ref{Sec:Otherstates} for  other model states). These states are the unique, highest density zero-mode wavefunctions of $(k+1)$-body pseudopotential Hamiltonians\cite{Simon:2007kx}.
Recent work\cite{Bernevig:2008aa} has shown the Read-Rezayi bosonic wavefunctions $\psi$ to be Jack polynomials $J^\alpha_{\lambda_0}$ indexed by a parameter $\alpha=-(k+1)$ and the densest-possible $(k,2)$-admissible `root' configuration\cite{stanley1989}:

\begin{align}
\label{eq:rootpartition}
n(\lambda_0)= \{k0k0k0\ldots k0k \}\, .
\end{align} For the $(k,2)$ states, the number of fluxes for the ground-state wavefunction is $N_\phi = 2(N/k-1)$.
 All the Jack polynomials at $\alpha=-(k+1)$ indexed by $(k,2)$-admissible root configurations are $(k,2)$-clustering polynomials, i.e. they vanish as $\prod_{i>k}(z-z_i)^2$ when $z=z_1=\ldots z_k$. 
 They form a basis for all many-body $(k,2)$-clustering polynomials and can be decomposed into a linear combination of Fock states with configurations squeezed from the root partition.

\section{The Orbital Entanglement Matrix (OEM)}
 \label{Sec:OEM}
 \subsection{Definition}
 
Consider dividing the set of single-particle orbitals $\{0,1,\ldots,N_\phi\}$ into two disjoint sets $A=\{0,1,\ldots l_A-1\}$ and $B=\{l_A,\ldots N_\phi\}$. As the single-particle orbitals are polynomially localized in the $\hat{\theta}$ direction, this partition in the single-particle momentum space roughly corresponds to an azimuthally symmetric spatial cut. 
  
The number of orbitals in $A$ ($B$) is $l_A$ ($l_B$), where $l_B=N_\phi+1-l_A$. Without loss of generality, 
 let $l_A\leq l_B$ ($l_A\geq l_B$ for $A$ and $B$ swapped).
Any occupation number state $|\lambda\rangle$ may be expressed as a tensor product $|\mu\rangle \otimes |\nu\rangle$ of states with partitions $\mu$ and $\nu$ belonging to the Hilbert spaces of $A$ and $B$ respectively. Thus, the model state can be decomposed as:
 \begin{equation}
 \label{eq:defnoem}
|\psi\rangle = \sum_\lambda b_\lambda |\lambda\rangle= \sum_{i,j} ({\bf C_f})_{ij} |\mu_i\rangle \otimes |\nu_j\rangle , 
\end{equation}
where the kets $\{|\mu_i\rangle\}$ and $\{|\nu_j\rangle\}$ form orthonormal bases that span the Hilbert spaces of $A$ and $B$. The matrix ${\bf C_f}$ is the full orbital entanglement matrix (OEM). 
The $(i,j)$th matrix element of the full OEM is equal to the coefficient of $|\mu_i +\nu_j\rangle$ in $|\psi\rangle$, i.e. $({\bf C_f})_{ij} = b_{\mu_i+\nu_j}$. 

In this article, we will almost exclusively deal with entanglement matrices. Unless stated otherwise, the rows (columns) of these matrices, for both the OEM defined in  Eq.\eqref{eq:defnoem} and for the PEM defined below, will be labeled by partitions $\mu_i$ ($\nu_j$) corresponding to the occupation basis states $|\mu_i\rangle$ ($|\nu_j\rangle$) in $A$ ($B$). 
The vector defined by the entries of a row/column in the entanglement matrix shall be referred to as row/column vector.

The OEM is explicitly constructed for the $4$-particle $\nu=1/2$ Laughlin state with an orbital cut after $3$ orbitals in Appendix~\ref{App:ExOEMLaughlin}.
 
 \subsection{Properties}
 \label{Sec:OEMProperties}
 ${\bf C_f}$ has a block-diagonal form; each block ${\bf C}$ is labelled by $N_A$, the number of particles in $A$, and $L_z^A$, the total $z$-angular momentum of the $N_A$ particles in $A$. Note that  $L_z^A=\sum_{i=1}^{N_A} \mu_i$ for the state $|\mu\rangle$, where $\mu_i$ here are the components of the partition $\mu$. Due to an unfortunate but necessary choice of notation,  $\mu_i$ also index the partitions of the Hilbert space of part $A$. In that case, $\mu_i$ is a partition by itself, and its components are $\mu_{i1}, \mu_{i2}, \ldots, \mu_{i N_A}$. The use of $\mu_i$ as a partition or as a component of a partition $\mu$ will be self-evident in the text.  To understand the origin of the block-diagonal structure of  ${\bf C_f}$, observe that $|\psi\rangle$ is an eigenstate of the particle-number operator $\hat{N}$ and the total $z$-angular momentum operator $\hat{L_z^{tot}}$. 
  As both operators are sums of one-body operators, $\hat{N}=\hat{N_A}\otimes\mathbb{I} + \mathbb{I}\otimes \hat{N_B}$ and $\hat{L_z^{tot}}=\hat{L_z^A}\otimes\mathbb{I} + \mathbb{I}\otimes \hat{L_z^B}$. Thus, every $|\lambda\rangle$ in Eq.\eqref{eq:defnoem} is labelled by the quantum numbers, $N$ and $L_z^{tot}$, while every $|\mu_i\rangle$ ($|\nu_j\rangle$) is labelled by $N_A$ ($N_B$) and $L_z^A$ ($L_z^B$). In the remainder of this article, the symbol ${\bf C}$ refers to the block of the full OEM ${\bf C_f}$ with labels $N_A$ and $L_z^A$. 

The reduced density matrices are  obtained from ${\bf C_f}$ as ${\boldsymbol \rho}_A={\bf C_fC_f}^\dagger$ and $\boldsymbol\rho_B={\bf C_f}^\dagger {\bf C_f}$. The block-diagonal structure of ${\bf C_f}$ carries over to the reduced density matrices and the rank of ${\boldsymbol \rho}_A$ and  ${\boldsymbol \rho}_B$ in each block is equal to that of $\bf C$. Neither ${\boldsymbol \rho}_A$ nor ${\boldsymbol \rho}_B$ uniquely determine all the coefficients of $|\psi\rangle$;  ${\bf C_f} $ clearly contains more information than either of the reduced density matrices.

The singular value decomposition of \C is given by:
\begin{equation}
\label{Eq:Cspecdecomp}
\sum_{i,j}{\bf C}_{ij} |\mu_i\rangle\otimes |\nu_j\rangle=\sum_{i=1}^{\textrm{rank}({\bf C})} e^{-\xi_i/2}|U_i\rangle\otimes | V_i\rangle.
\end{equation}
The kets on the left-hand-side of Eq.~\eqref{Eq:Cspecdecomp} are defined as in Eq.~\eqref{eq:defnoem}. $|U_i\rangle$ and $|V_i\rangle$ are the singular vectors in the Hilbert spaces of $A$ and $B$ restricted to a fixed particle number and $z$-angular momentum. 
  They are linear combinations of the occupation number basis vectors $|\mu_i\rangle$ and $|\nu_j\rangle$. The $\xi_i$'s are the `energies'  plotted as a function of $L_z^A$ in the orbital entanglement spectrum (OES) introduced in [\onlinecite{Li:2008aa}]. A typical OES is shown in Fig.~\ref{fig:oem_n_9_2s_16_na_4_la_8} for the $9$ particle $1/2$ Laughlin state with orbital cut $l_A=8$ in the sector $N_A=4$. Note the counting of the entanglement levels, $\{1,1,2,3,5\ldots\}$, starting from the right edge of the spectrum.

The number of finite energies (rank($\bf C$)) at each $L_z^A$ is independent of the geometry of the 2-d surface and the symmetrization factors arising due to multiple particles occupying the same orbital. Let ${\bf C_f}^{d}$ and ${\bf C_f}^{s}$ be the full OEMs in the disc and sphere geometry, or in any other two genus $0$ geometries. Modifying the geometry of the surface changes the normalization of the single-particle orbitals (the quantum mechanical normalization); thus every $b_\lambda$ in the expansion of $|\psi\rangle$ in Eq.~\eqref{eq:defnoem} in the disc basis is multiplied by a factor $\mathcal{N}(\lambda)=\prod_{i=1}^N \mathcal{N}(\lambda_i)$. when expanded in the single-particle  orbital basis on the sphere. $\mathcal{N}(j)$ is a factor relating the normalization of orbital $j$ on the disc to that on the sphere. The OEM's on the disc
  and the sphere are thus related as:
\begin{eqnarray}
 \label{Eq:relngeometry}
|\psi\rangle &=& \sum_{i,j} ({\bf C_f}^d)_{ij} |\mu_i^{d}\rangle \otimes |\nu_j^{d}\rangle   \nonumber \\
&=& \sum_{i,j} ({\bf C_f}^{d})_{ij} \mathcal{N}(\mu_i^{d}) \mathcal{N}(\nu_j^{d}) |\mu_i^{s}\rangle \otimes |\nu_j^{s}\rangle ,
\end{eqnarray}
where the superscripts $d$ and $s$ refer to the disc and sphere geometries, or to any other two genus $0$ geometries. 
${\bf C_f}^{s}$ is obtained from ${\bf C_f}^{d}$ by multiplying whole rows and columns by normalization factors; thus $\textrm{rank}({\bf C_f}^{s}) = \textrm{rank} ( {\bf C_f}^{d})$. An identical argument shows the rank of ${\bf C_f}$ to be independent of the symmetrization factors that arise in the normalization of the \emph{many-body} states constructed from normalized single-particle orbitals. We are therefore free to work in an unnormalized single-particle basis from this point.

 For a given cut $l_A$ in orbital space, the \emph{maximum} number of particles that can form a $(k,2)$-clustering droplet in $A$ is defined to be the natural number of particles $N_{A,nat}$. Explicitly, $N_{A,nat}=k\lfloor (l_A+1)/2\rfloor$  where $\lfloor x\rfloor$ is the integer part of $x$. The OES at $N_{A,nat}$ is called the natural spectrum. In the thermodynamic limit, the counting of the OES is  conjectured to be identical to the counting of the modes of the CFT describing the edge for values $l_A, N_A\rightarrow \infty$ such that $l_A/N_\phi\rightarrow$  const.$(>0)$ and $N_A/N_{A,nat}\rightarrow 1$.

For future reference, $L_{z,min}^A$ denotes the minimum $z$-angular momentum of the $N_A$ particles in $A$:
\begin{align}\label{eq:lzmin}
L^A_{z,min}=\lfloor{N_A/k}\rfloor(2 N_A-k \lfloor{N_A/k}\rfloor-k). 
\end{align}
 We stress that $L_{z,min}^A$ is the \emph{maximum} value on the $x$-axis of the \emph{numerically} generated entanglement spectra existing in the literature, due to the different indexing scheme in the text and the figures (see also the discussion in Section~\ref{Sec:Notation}).  For instance, in Fig.~\ref{fig:oem_n_9_2s_16_na_4_la_8}, $L^A_{z,min}$ describes the sector of the OES at $L_z^A=20$.

\section{The Particle Entanglement  Matrix (PEM)}
\label{sec:pemintro}
\subsection{Definition}
\label{sec:pemdef}
In the orbital cut that we just discussed, the Hilbert space of $A$ at some $(N_A, L_z^A)$ was spanned by the possible occupation configurations $|\mu\rangle$ of $N_A$ particles, such that $\sum_{i=1}^{N_A} \mu_i = L_z^A$ and $\mu_i<l_A \,\forall i$. 
We now consider making a cut of a FQH state $\ket{\psi}$ in particle space by dividing the $N$ particles into groups $A$ and $B$ with $N_A$ and $N_B=N-N_A$ particles. Let $N_A\leq N_B$. The Hilbert space of $A$($B$) is now spanned by all possible occupation configurations of $N_A$($N_B$) particles \emph{in the full single-particle orbital basis} of the state $\ket{\psi}$ and contains the smaller Hilbert space of $A$($B$) with the orbital restriction. Just as in the previous section, we can write the model wavefunction $|\psi\rangle$ as:
 \begin{equation}
 \label{eq:defnpem}
|\psi\rangle = \sum_\lambda b_\lambda |\lambda\rangle= \sum_{i,j} ({\bf P_f})_{ij} |\mu_i\rangle \otimes |\nu_j\rangle ,
\end{equation}
where the kets $\{|\mu_i\rangle\}$ and $\{|\nu_j\rangle\}$ are Fock states of $N_A$ and $N_B$ particles in the full single-particle  orbital space. The matrix ${\bf P}_f$ is the full particle entanglement matrix (PEM). 
As was the case for the OEM, the matrix elements of the PEM are directly related to the weights of the model wavefunction by $({\bf P}_f)_{ij}=b_{\mu_i+\nu_j}$. 

\subsection{Properties}
For a given cut with $N_A$ particles in $A$, ${\bf P}_f$ is block-diagonal in the angular momentum of part $A$, $L_z^A$. The block of ${\bf P}_f$ at fixed $(L_z^A, N_A)$ shall be denoted by ${\bf P}$. 
The reduced density matrices of part $A$ and  $B$ are given by ${\boldsymbol \rho}_A={\bf P}_f {\bf P}^\dagger_f$  and ${\boldsymbol \rho}_B={\bf P}^\dagger_f {\bf P}_f$ respectively. They are block-diagonal in $L_z^A$ and have the same rank as ${\bf P}_f$ in each block. 
In the same spirit as the discussion of the OEM, we define the singular value decomposition of the PEM by:
\begin{align}
\label{Eq:PSingDecomp}
\sum_{i,j} ({\bf P})_{ij} |\mu_i\rangle \otimes |\nu_j\rangle&=\sum_{i} e^{-\xi_i/2} |U_i\rangle \otimes |V_i\rangle,
\end{align}
where the singular vectors $|U_i\rangle$ and $|V_i\rangle$ are orthonormal vectors in the Hilbert spaces of $A$ and $B$ restricted to fixed angular momentum. The plot of the `energies' $\xi_i$ vs $L_z^A$ is called the particle entanglement spectrum (PES)\cite{sterdyniak:2010aa}. In Fig.~\ref{fig:pem_n_9_2s_16_na_4_la_8} we show the PES of the $9$ particle $1/2$ Laughlin state for the particle cut $N_A=4$. Note the \emph{same} entanglement level counting starting from the right edge of the spectrum,  $(1,1,2,3,5\ldots)$, in the OES in Fig.~\ref{fig:oem_n_9_2s_16_na_4_la_8} and the PES in Fig.~\ref{fig:pem_n_9_2s_16_na_4_la_8}.

In the spherical geometry, the PEM is labelled by an additional quantum number as compared to the OEM\cite{sterdyniak:2010aa}--- the total angular momentum of $A$, $(\vec L^A)^2$. 
Consequently, the eigenvalues of the block of the reduced density matrix with $(\vec L^A)^2=\ell(\ell+1)$ have $(2\ell+1)$-fold degeneracy. This multiplet structure, apparent in the PES in Fig.~\ref{fig:pem_n_9_2s_16_na_4_la_8}, does not play any role in our discussions about the counting of the PES in this article. 

 The OEM ${\bf C} $ at fixed particle number $N_A$ and angular momentum $L_z^A$ is a sub-matrix of the larger PEM, ${\bf P}$, at the same angular momentum. $L_{z,min}^A$ is the same for both cuts.

\subsection{Rank}
\label{Sec:Ptilde}
The property that defines the $k$-clustered model state $\psi(z_1,\ldots, z_N)$ uniquely is that it is the lowest degree symmetric polynomial that vanishes when $(k+1)$ particles are at the same position. Similar clustering conditions characterize every ground-state of a pseudopotential Hamiltonian. This vanishing property must persist when we divide the particles into two groups and re-write the model state in Eq.~\eqref{eq:defnpem} as: 

\begin{multline}
\label{Eq:Pfullsingular}
\psi(z_1,\ldots,z_N)=\\ \sum_{L_z^A} \sum_{i} e^{-\xi_i/2}  \, \langle z_1, \ldots, z_{N_A}|U_i\rangle \otimes \langle z_{N_A+1},\ldots,z_N |V_i\rangle,
\end{multline}
 using Eq.~\eqref{Eq:PSingDecomp} at each $L_z^A$. If we choose $(k+1)$ particles in group $A$, say $z_1,\ldots,z_{k+1}$, to be at the same position $z$, then the state must vanish at every $L_z^A$. Further, as the singular vectors in $B$ form an orthonormal basis:
\begin{multline}
\label{Eq:psivanishing}
\psi(z,\ldots,z,z_{k+2},\ldots, z_N)=0\\
\Rightarrow   e^{-\xi_i/2} \langle z, \ldots, z, z_{k+2},\ldots,z_{N_A}|U_i\rangle =0, \, \forall i,L_z^A\, .
\end{multline}
 A similar relation holds when $A$ and $B$ are interchanged. 
We conclude that the singular vectors, $\langle z_1, \ldots, z_{N_A} | U_i\rangle$ and $\langle z_{N_A+1},\ldots,z_N |  V_i\rangle$, must also be clustering polynomials that vanish when $(k+1)$ particles are at the same position. 
A basis for clustering polynomials is the set of Jack polynomials, $J^{\alpha}_{\tilde \mu}$, indexed by $\alpha=-(k+1)$ and the $(k,2)$-admissible partition $\tilde{\mu}$\cite{stanley1989, Bernevig:2008aa, Bernevig:2008kx}. 
$\psi$ can therefore be expanded in the Jack basis as:
\begin{multline}
\label{Eq:psiexpansionJack}
\psi(z_1,\ldots,z_N)=\\ \sum_{i,j} ({\bf M}_f)_{ij} J^\alpha_{\tilde \mu_i}(z_1,\ldots,z_{N_A}) J^\alpha_{\tilde \nu_j}(z_{N_A+1},\ldots,z_{N}) ,
\end{multline}
where $\tilde\mu_i$ and $\tilde \nu_j$ denote $(k,2)$-admissible partitions of $N_A$ and $N_B$ particles respectively.  
The matrix ${\bf M}_f$ is block-diagonal in angular momentum $L_z^A$; let ${\bf M}$ refer to the block of $\mathbf{M_f}$ at fixed value of $L_z^A$. The row and column dimensions of ${\bf M}$ are much smaller than those of ${\bf P}$ because the $(k,2)$-admissible partitions of $N_A$ and $N_B$ form a small subset of the set of all partitions of  with fixed $L_z^A$ and $L_z^B$ respectively. Nevertheless, as Eq.~\eqref{Eq:Pfullsingular} and \eqref{Eq:psiexpansionJack} are equal, $\bf M$ and $\bf P$ must have the same rank. Thus, the minimum of the row and column dimension of ${\bf M}$ bounds the rank of the PEM from above. 

 Let us reformulate what we have just shown in a more familiar language and argue for the saturation of the bound. 
The smaller dimension of $\bf M$ is  the row dimension because $N_A\leq N_B$. It is equal to the number of distinct bulk quasi-hole excitations of the $(k,2)$-clustering model state of $N_A$ particles at angular momentum $L_z^A$ on a sphere pierced by the number of fluxes of the original state, $N_\phi = 2/k(N-k)$. 
For $L_z^A\leq L_{z,min}^A + \lfloor N_A/k\rfloor$, this is the number of distinct bulk quasi-hole excitations in the thermodynamic limit, i.e. no finite-size effects occur.
We have thus bounded the level counting of the PES by the number of bulk quasihole excitations at each $L_z^A$. 
Without further symmetry-induced constraints on the reduced density matrices (we have already used all the symmetries available in the state), we expect this bound to be saturated. 
In the thermodynamic limit  ($N_A,N\rightarrow \infty$ such that $N_A/N>0$), we therefore argue that the level counting of the entire PES is identical to the number of the bulk quasi-hole excitations. This bound saturation can be proved exactly for the Laughlin states\cite{benoitestienneprivatecommunication}.

It is beneficial to identify a set of rows and columns in ${\bf P}$ with the same rank as the full matrix. 
Consider the rows and columns labelled by  the $(k,2)$-admissible partitions. 
This sub-matrix of ${\bf P}$ is denoted by ${\bf \tilde P}$ and has the same dimensions as ${\bf M}$.  In Appendix~\ref{App:tildePMrank},  we show that  ${\bf \tilde P}$ and ${\bf M}$ have the same rank. 
 $\bf\tilde {P}$ will play a prominent role in the proof establishing the bulk-edge correspondence in the entanglement spectra.

\begin{figure}[htbp]
  \begin{center}
    \includegraphics[width=0.4\textwidth]{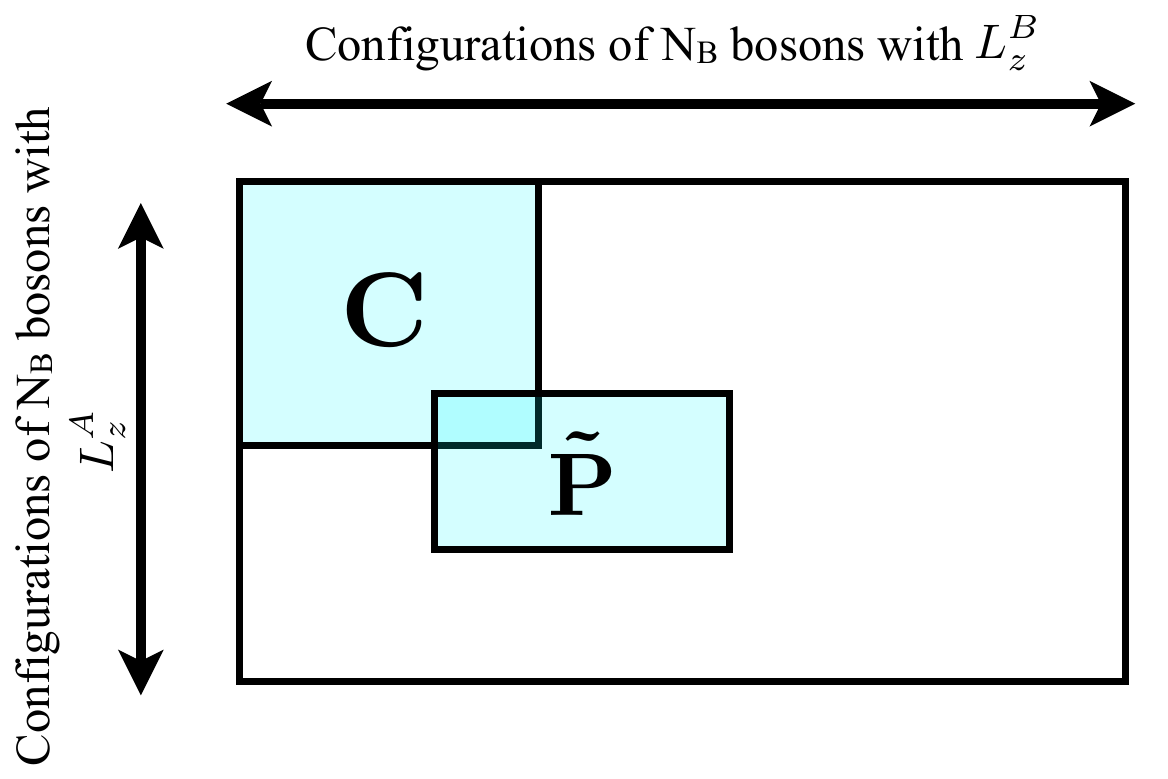}
    \caption{A cartoon of the various sub-matrices in ${\bf P}$}
    \label{fig:pemcolsrows}
  \end{center}
\end{figure}

\section{Clustering Constraints}

\label{Sec:Squeezingconstraints}

In this section, we introduce the $(k+1)$-body clustering constraints that relate the rank of the PEM and the OEM of the clustering model states and  establish the bulk-edge correspondence in the entanglement spectra. 
The Read-Rezayi model wavefunctions $\psi_{(k,2)}(z_1,z_2\ldots z_N)$ are single Jack polynomials labelled by a root partition $\lambda_0$ (Eq.~\eqref{eq:rootpartition}), and a parameter $\alpha=-(k+1)$. 
They satisfy $(k,2)$-clustering --- they are non-zero when a cluster of $k$ particles is at the same point in space $z=z_1=z_2=\ldots z_k$, but vanish as the second power of the distance between the $(k+1)$st particle and the cluster as $z_{k+1}\rightarrow z$. The clustering property imposes a rich structure on $\psi_{(k,2)}(z_1,z_2,\ldots, z_N)$.
 All the partitions $\lambda$ that arise in the expansion of $|\psi\rangle$ in the many-body occupation basis ($|\psi\rangle=\sum_\lambda b_\lambda |\lambda\rangle$) are dominated by $\lambda_0$. Furthermore, all the coefficients $b_\lambda$ are known up to a multiplicative constant. In the Jacks, this constant is chosen so that $b_{\lambda_0}=1$. 
In other words, the clustering property and the requirement to be the densest possible wavefunction determine $\psi_{(k,2)}(z_1,z_2,\ldots, z_N)$ uniquely up to an overall normalization constant. Here, we formulate the conditions imposed by clustering on $\psi_{(k,2)}(z_1,\ldots, z_N)$ as linear, homogeneous equations on the coefficients $b_\lambda$.

 \subsection{Derivation}\label{Sec:ClusteringDerivation}
Let us introduce a `deletion' operator $d_i$ for orbital $i$ such that:
\begin{equation}
\label{eq:delop}
d_i |\lambda\rangle = \left\{\begin{array}{cc}
 0& , i\notin\lambda \\
| \lambda\backslash\{i\}\rangle&, i\in\lambda
 \end{array}\right. 
\end{equation}
$\lambda\bs\{ i\}$ is the partition with a single occurrence of the orbital $i$ removed from it. The `deletion' operators commute with each other. In Appendix~\ref{App:PseudoHam}, we derive the relation between these operators and the annihilation operators in the normalized single-particle basis.

We now separate the coordinates of $k+1$ particles from the rest and rewrite $\psi_{(k,2)}(z_1,z_2,\ldots, z_N)$ as:
\begin{multline}
\label{Eq:psi_k_plus_one_coords}
\psi_{(k,2)}(z_1,\ldots, z_N)=\\
\sum_{l_1\ldots, l_{k+1}=0}^{N_\phi} \left(\prod_{j=1}^{k+1} z_j^{l_j}\right) \langle z_{k+2},\ldots, z_N |\prod_{j=1}^{k+1} d_{l_j} |\psi\rangle\, ,
\end{multline} 
and form a cluster by bringing the $k$ particles with coordinates $z_1,\ldots, z_k$ to the same position $z$. When $z_{k+1}=z$, the LHS vanishes and Eq.\eqref{Eq:psi_k_plus_one_coords} becomes:
\begin{equation}
0= \sum_{l_1,\ldots, l_{k+1}=0}^{N_\phi} z^{\sum_{j=1}^{k+1}l_j} \langle z_{k+2},\ldots, z_N |\prod_{i=1}^{k+1} d_{l_i} |\psi\rangle .
\end{equation} 
The right-hand-side is a polynomial in an arbitrary complex number $z$, and has to vanish for every power $\beta=\sum_{j=1}^{k+1}l_{j}$ of $z$ to satisfy the above equation. Thus, the constraints on $|\psi\rangle$ are:
\begin{equation}\label{eq:clustering}
\left(\sum_{l_1,\ldots l_{k}=0}^{N_\phi} d_{\beta-\sum_{j=1}^k l_j} \prod_{j=1}^{k}d_{l_j}\right)|\psi\rangle =D_\beta|\psi\rangle= 0.
\end{equation}
$\beta$ is the $z$-angular momentum of $(k+1)$-particles; it ranges from $0$ to $N_\phi(k+1)$. The equation above requires any clustering wavefunction $|\psi\rangle$ to be simultaneously annihilated by the destruction operators $\{D_i\,,i=0\ldots N_\phi(k+1)\}$.

\subsection{Properties}
Every value of $\beta$ in Eq.~\eqref{eq:clustering} yields,  in general, a large number of linear relations between the coefficients of $|\psi\rangle$.  Let $S_\beta$ be the set of all partitions of $N$ particles such that the sum of the $z$-angular momentum of $(k+1)$ particles is $\beta$. For \emph{every} occupation configuration of $N-(k+1)$ particles, Eq.~\eqref{eq:clustering} relates the coefficients of partitions $\lambda\in S_\beta$ in the expansion of $|\psi\rangle$. Examples of such relations are given in Appendix~\ref{App:ClusteringConstraintsExamples}.

The set of linear, homogeneous equations in Eq.\eqref{eq:clustering} are linearly dependent. The dimension of the null-space of the set is \emph{exactly} one for the densest possible wavefunction, i.e. the vector of coefficients $\{b_\lambda\}$ is uniquely determined up to an overall multiplicative factor. Since the solution to Eq.\eqref{eq:clustering} causes $\psi$ to vanish when any cluster of size greater than $k$ is formed in real space, we conclude that the set in Eq.\eqref{eq:clustering} includes \emph{all} constraints imposed on $\psi(z_1,\ldots, z_N)$ due to clustering.

Equivalently, we are describing model FQH wavefunctions that are the unique, highest density zero-modes of the Haldane pseudopotentials or their generalization to the $k+1$-body interaction [\onlinecite{Simon:2007kx}].
In fact, the destruction operators above are the fundamental clustering operators from which the Haldane pseudopotentials can be obtained as the translationally-invariant supersymmetric form:

\begin{equation}
\label{Eq:Pseudoham}
 H =\sum_\beta f(\beta) D_\beta^\dagger D_\beta.
\end{equation}
$f(\beta)$ can be derived at each $k$; in Appendix~\ref{App:PseudoHam}, we work through the $k=1$ case.


\section{Relating the OES and PES counting}
\label{Sec:Theproof}
 We now have all the ingredients necessary to relate the counting of the PES to that of the OES for a given number of particles in $A$ and cut in orbital space. In Sec.~\ref{Sec:Ptilde}, we constructed the sub-matrix $\bf\tilde{P}$ of $\bf P$, labelled by $(k,2)$-admissible configurations with the same rank as $\bf P$. We now use the clustering relations derived in the previous section to express the row/column vectors of $\bf\tilde{P}$ in terms of those of the OEM $\bf C$. When possible in finite-size, i.e. up to a certain value of $L_z^A$, this procedure proves that the PES and the OES have the same counting. In the thermodynamic limit, this procedure establishes the equality of the level counting of the \emph{entire} PES to  OES when $N_A/N_{A,nat} \rightarrow 1 $, thus proving a significant part of the Li-Haldane conjecture. 
 
 The argument below applies equally well to row and column vectors.  To keep the discussion concise,  we formulate it using row vectors alone.

\subsection{Systemizing the constraints}
The biggest challenge in relating the row vectors of $\bf{\tilde{P}}$ to those in the OEM lies in identifying a set of linearly independent equations in the entire set of clustering constraints. To this end, we introduce a few quantities characterizing a partition $\mu$. $n_m(\mu)$ below refers to the occupation number of the $m$th orbital in partition $\mu$. The orbital cut is after $l_A$ orbitals.

\emph{The unit cell---}  We divide the single-particle  orbital space such that the $j$th unit cell contains the orbitals of $z$-angular momentum $2j$ and $2j+1$, and $j\in[0,\ldots, N_\phi/2)$. {\bf As} the total number of single-particle orbitals is odd for the bosonic $(k,2)$-clustering states, the orbital with angular momentum $N_\phi$ is its own unit cell with index $N_\phi/2$. Every orbital  belongs to exactly one unit cell.

\emph{The intact unit cell---} 
The $j$th unit cell of a partition $\mu$ is said to be intact if the occupation numbers of the orbitals with angular momentum $0,\ldots,2j+1$ are identical to those in the root configuration Eq.~\eqref{eq:rootpartition}, i.e. if $n_i(\mu)=n_i(\lambda_0)$ for $i=0,...,2j+1$. Clearly, the $j$th unit cell can only be intact if all unit cells $0,\ldots, j-1$ are intact. 

\emph{The number of intact unit cells in part $A$--- } 
The number of intact unit cells in part $A$, $\Delta_\mu$, is the number of intact unit cells to the left of the orbital cut in $n(\mu)$.

\begin{figure}[htbp]
  \begin{center}
    \includegraphics[width=0.4\textwidth]{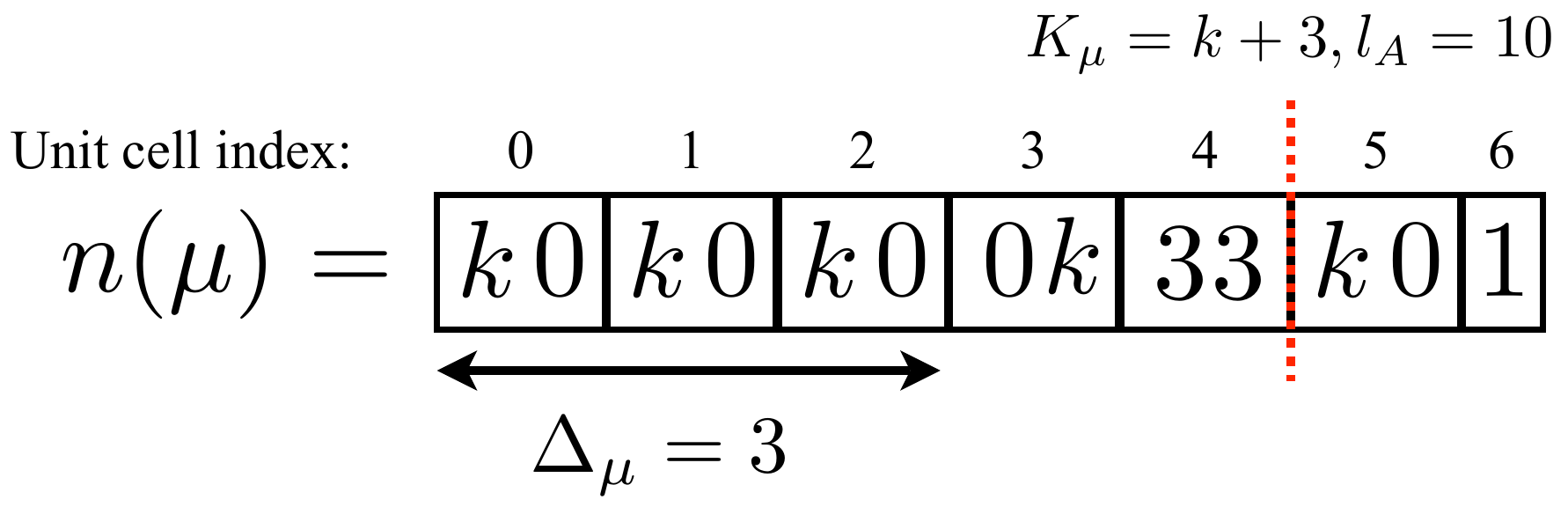}
     \includegraphics[width=0.4\textwidth]{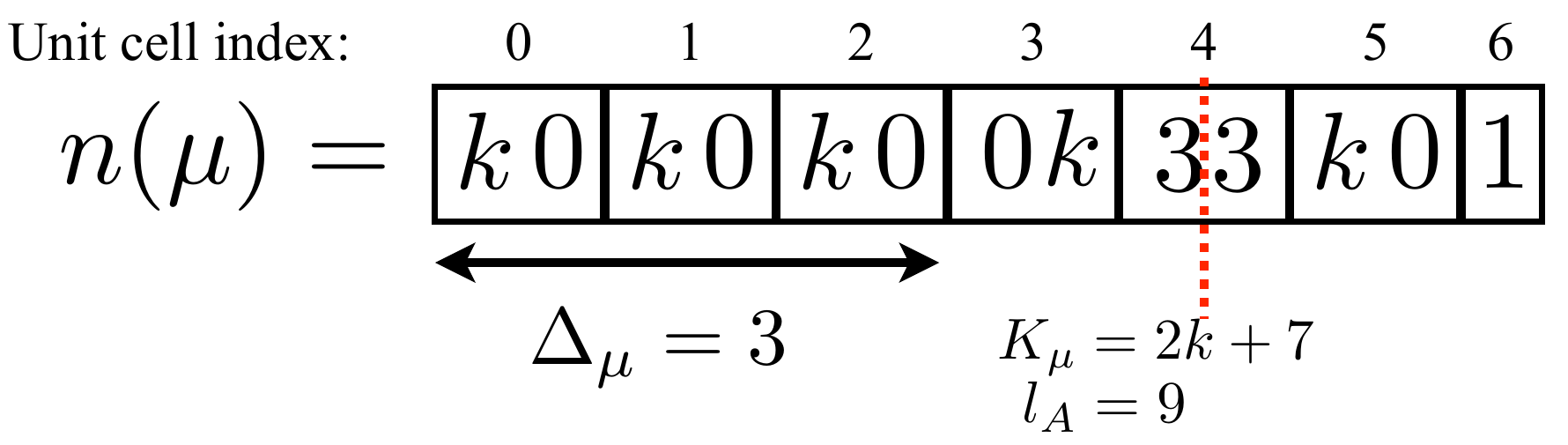}
    \caption{The occupation configuration of a generic partition $\mu$ with the unit cells, the number of intact unit cells $\Delta_\mu$ and the distance from cuts after $l_A=9$ (top) and $l_A=10$ (bottom) shown. $N_\phi=12$ here. }
    \label{fig:dist_intactunitcells}
  \end{center}
\end{figure}

\emph{Distance from the cut--- } If we were to number the orbitals to the right of the cut as $1,2,\ldots$, then the distance from the cut is defined as the sum of the indices of the occupied orbitals to the right of the orbital cut in $n(\mu)$. The distance from the cut, $K_\mu$, is given by:
\begin{align}
K_\mu=\sum_{m=l_A}^{N_\phi} n_m(\mu)(m - l_A + 1).
\end{align}
 $K(\mu)=0$ for a partition $\mu$ labeling a row of the OEM; for a general partition, it represents the distance in orbital units that all the particles to the right of the cut need to traverse to cross the cut. In Fig.~\ref{fig:dist_intactunitcells}, we pick as an example a generic partition $\mu$ and identify the number of intact unit cells in $A$, $\Delta_\mu$,  and the distance from the cut, $K_\mu$, for two different orbital cuts.

\subsection{The method}
\label{Sec:Thegeneralmethod}

Consider the blocks $\bf\tilde{P}$ and $\bf C$ at $N_A$ and $L_z^A=L_{z,min}^A+\ell$. We can relate all row vectors of $\mathbf{\tilde P}$ to row vectors in the OEM if:
\begin{align}
\Delta_{\tilde \mu_0} \geq K_{\tilde \mu_0} ,
\end{align}
where $n(\tilde\mu_0)$ is a $(k,2)$-admissible occupation configuration and labels a row of $\bf \tilde{P}$:
\begin{align}
 \label{eq:tildemu0}
 n(\tilde{\mu_0})&= \{\underbrace{k0\ldots k0}_{2\lfloor (N_A-1)/k\rfloor } x\underbrace{0\ldots0}_{\ell-1 }10\ldots 0\}.
\end{align}
The value of $x$ is fixed by the total particle number being $N_A$ ($x=(N_A-1)-k\lfloor (N_A-1)/k \rfloor$).  $\tilde\mu_0$ dominates all the other partitions that label rows of the PEM. It has the maximum (total) number of intact unit cells possible, $\lfloor (N_A-1)/k \rfloor$.
The constraint, $ \Delta_{\tilde\mu_0}\geq K_{\tilde\mu_0}$, sets an upper bound on the values of $\ell$ for which we can express all row vectors in $\bf\tilde{P}$ in terms of those in $\bf C$: 
\begin{widetext}
\begin{align}\label{eq:ellmax}
\ell\leq \left\{
\begin{array}{cl}
\Delta_{\tilde\mu_0}-k\bar\Delta_{\tilde\mu_0}^2-(2\bar\Delta_{\tilde\mu_0}+1)(x+1) & \, \mbox{for $l_A$ even, }l_A\leq 2\lfloor (N_A-1)/k\rfloor \\
\Delta_{\tilde\mu_0}-k\bar\Delta_{\tilde\mu_0}(\bar\Delta_{\tilde\mu_0}-1)-2(x+1)\bar\Delta_{\tilde\mu_0}&\, \mbox{for $l_A$ odd, }l_A\leq 2\lfloor (N_A-1)/k\rfloor\\
l_A-\Delta_{\tilde \mu_0}+1 &\, \mbox{for }  l_A>2\lfloor (N_A-1)/k\rfloor\, ,
\end{array}\right.\,
\end{align}
\end{widetext}
where we abbreviated the difference of the total number of intact unit cells and those only in $A$ by
$\bar\Delta_{\tilde\mu_0}=\lfloor (N_A-1)/k\rfloor-\Delta_{\tilde\mu_0}$. 
 For such values of $\ell$, the method of proof can be broken into two steps ---
\begin{enumerate}[I]

\item  \label{Item:Result2} 
If $ \Delta_{\tilde\mu_0}\geq K_{\tilde \mu_0}$, then $ \Delta_{\tilde\mu}\geq K_{\tilde\mu}$ for all $(k,2)$-admissible partitions $\tilde\mu<\tilde\mu_0$. 

\item \label{Item:Result1} 
If $\Delta_\mu\geq K_\mu$ for a partition $\mu$, then the row vector labelled by $\mu$ in $\bf P$ can be expressed as a linear combination of row vectors in the OEM $\bf C$ alone.
\end{enumerate}
 We prove these statements rigorously in the Appendices~\ref{App:KDeltaofkr} and \ref{app:step1}. The first step shows that the $\Delta_{\tilde\mu}\geq K_{\tilde\mu}$ for all partitions $\tilde{\mu}$ labeling rows of $\tilde{P}$; the second assures that all these rows can be written as linear combinations of rows in the OEM alone.

An identical argument can be repeated for the columns; let $\tilde\nu_0$ be the $(k,2)$-admissible partition that dominates all partitions of $L_z^B$ into $N_B$ parts. For values of $\ell$ such that $ \Delta_{\tilde\mu_0}\geq K_{\tilde \mu_0}$ and $ \Delta_{\tilde\nu_0}\geq K_{\tilde \nu_0}$, the OEM and PEM have the same counting in finite-size. 
$\ell$ is usually of order $N_A/k$. We can therefore analytically prove the bulk-edge correspondence in the entanglement spectra for a few values of $L_z^A$ near $L_{z,min}^A$ (the right edge of the numerically generated spectra). 
Note that at these values of $\ell$, it was proved in [\onlinecite{sterdyniak:2010aa}]  that the counting of the PES is bounded from above by the number of bulk quasi-hole excitations in the thermodynamic limit. We argued in Sec.~\ref{Sec:Ptilde} that due to the absence of extra-symmetries, this bound should be saturated.

The heart of the proof lies in the use of the $(k+1)$-clustering condition at the $z$-angular momentum of the $k$ particles in the right-most intact unit cell in part $A$ and one particle occupying an orbital to the right of the cut. This relates a \emph{single} row vector belonging to the PEM with $\Delta_\mu$ and $K_\mu$ to row vectors with $\Delta_{\mu'}= \Delta_\mu-1$ and $K_{\mu'} \leq K_\mu -1$. The clustering conditions thus allow us to replace a row vector whose partition has distance $K_\mu$ with a linear combination of row vectors whose partitions have distances reduced by \emph{at least} one at the cost of using a \emph{single} intact unit cell. If $\Delta_\mu\geq K_\mu$ for a partition $\mu$, then iterating this procedure provides a linear relation between the row vector labelled by partition $\mu$ and row vectors with distance zero, i.e. row vectors of the OEM $\bf C$. 

 In the thermodynamic limit, $N_A$ and $l_A$ scale with $N$. Consequently, the number of intact unit cells in $A$ ($B$) in $\tilde \mu_0$  ($\tilde \nu_0$) denoted by $\Delta_{\tilde{\mu}_0}$ ($\Delta_{\tilde{\nu}_0}$) scales with $N$. 
For the upper bound on $\ell$ in Eq.~\eqref{eq:ellmax} to scale with $N$, $\bar\Delta_{\tilde\mu_0}$ and  $\bar\Delta_{\tilde\nu_0} $ have to grow \emph{slower} than $\sqrt{N}$:
\begin{eqnarray*}
\Delta_{\tilde{\mu_0}}, \Delta_{\tilde{\nu_0}} \sim N \\
\bar\Delta_{\tilde{\mu_0}}, \bar\Delta_{\tilde{\nu_0}} \sim N^g \, ,\, \, g<1/2\, .
\end{eqnarray*}
As:
\begin{equation*}
\bar\Delta_{\tilde{\mu_0}}\sim|N_A-N_{A,nat}| \, , \,  \bar\Delta_{\tilde{\nu_0}}\sim|N_B-N_{B,nat}| \, ,
\end{equation*}
 $|N_A-N_{A,nat}|$ must grow \emph{slower} than $\sqrt{N}$. Thus, in the thermodynamic limit, condition \eqref{eq:ellmax} can only be fulfilled if $N_A/N_{A,nat}\rightarrow1$. In OES sectors such that $N_A/N_{A,nat}\rightarrow1$, the level counting of the OES is then identical to the level counting of the PES at $N_A=N_{A,nat}$ for \emph{all} angular momenta $L_z^A$, thus proving the bulk-edge correspondence in the entanglement spectra.

\subsection{Illustrative examples}

 The proof of the full method is presented in the appendices; here we illustrate the more formal ideas with examples of the general method at work for the $k=1,2$ wavefunctions.

\subsubsection{At $k=1$:}
Consider the $\nu=1/2$ Laughlin state of $N=7$ bosons with $N_\phi=12$ and $L_z^{tot}=42$. 
Let $l_A=6$ and the number of particles in $A$ be the natural number $N_A=N_{A,nat}=3$. 
We consider the entanglement level counting of the OES and the PES at $L_z^A=L_{z,min}+\ell=L_{z,min}+3$. 
We first verify that the conditions, $ \Delta_{\tilde\mu_0}\geq K_{\tilde \mu_0}$ and $ \Delta_{\tilde\nu_0}\geq K_{\tilde \nu_0}$, are satisfied.
The occupation configurations of $\tilde{\mu_0}$ and $\tilde{\nu_0}$ are:
\begin{eqnarray*}
n(\tilde{\mu_0}) &=&\{ 101000 \,|\,0100000 \} \qquad K_{\tilde{\mu_0}}=2, \Delta_{\tilde{\mu_0}}=2 \\
n(\tilde{\nu_0}) &=& \{000100\,|\,0010101\}  \qquad K_{\tilde{\nu_0}}=3, \Delta_{\tilde{\nu_0}}=3
\end{eqnarray*}
The cut in orbital space is indicated in the occupation configurations by the `$|$' symbol. 
Hence, the method discussed in the previous section should prove the equality of the ranks of the OEM and the PEM at this $L_z^A$. 

The occupation configurations of the $(1,2)$-admissible partitions labeling the rows of $\bf\tilde{P}$ are:
\begin{eqnarray}
\label{Eq:Laugh_Na_3_Lz_3}
n(\tilde{\mu_0}) &=& \{1 0 1 0 0 0 \, | \, 0 1 0 \ldots 0\} \qquad K_{\tilde{\mu_0}}=2, \Delta_{\tilde{\mu_0}}=2 \nonumber\\
n(\tilde{\mu_1}) &=& \{1 0 0 1 0 0 \, | \, 1 0 0 \ldots 0\} \qquad K_{\tilde{\mu_1}}=1, \Delta_{\tilde{\mu_1}}=1 \nonumber\\
n(\tilde{\mu_2}) &=& \{0 1 0 1 0 1 \, | \, 0 0 0 \ldots 0\} \qquad K_{\tilde{\mu_2}}=0, \Delta_{\tilde{\mu_2}}=0\, . \nonumber\\
\end{eqnarray}
$\tilde{\mu_2}$ labels a row that already belongs to the OEM $\bf C$. 
We now relate the row labeled by the partition $\tilde\mu_1$ to rows of the OEM. 
In $n(\tilde{\mu_1})$, only the $0$th unit cell is intact and the particle to the right of the cut occupies the orbital with index $6$.  
Let us pick the 2-body clustering constraint at $\beta=6$  (the sum of the $z$-angular momentA of the particle in the intact unit cell and the particle to the right of the cut) in Eq.~\eqref{eq:clustering}:
\begin{equation}
\label{Eq:Laugh_beta6}
(2(d_0d_6 + d_1d_5 + d_2d_4) + d_3d_3) |\psi\rangle  = 0
\end{equation}
For every occupation number configuration of $(N-2)$ bosons with angular momentum $(L_z^{tot}-\beta)$, Eq~\eqref{Eq:Laugh_beta6} gives one linear relation. 
The appropriate occupation number configuration for our purpose is $n([3]+\nu_j)$, as: 
\begin{align}
d_0d_6\left(|\tilde\mu_1+\nu_j\rangle\right)=|[3]+\nu_j\rangle \, .
\end{align}
The partitions $\nu_j$ of $L_z^B$ into $N_B=4$ parts label the columns of the PEM $\bf P$. Eq.~\eqref{Eq:Laugh_beta6} then relates the row indexed by $\tilde{\mu_1}$ to row vectors indexed by following partitions:

\begin{eqnarray}
n({\mu_1}) &=& \{0 1 0 1 0 1 \, | \, 0 0 0 \ldots 0\} \qquad K_{{\mu_1}}=0, \Delta_{{\mu_1}}=0 \nonumber\\
n({\mu_2}) &=& \{0 0 1 1 1 0 \, | \, 0 0 0 \ldots 0\} \qquad K_{{\mu_2}}=0, \Delta_{{\mu_2}}=0 \nonumber\\
n({\mu_3}) &=& \{0 0 0 3 0 0 \, | \, 0 0 0 \ldots 0\} \qquad K_{{\mu_3}}=0, \Delta_{{\mu_3}}=0 \, .\nonumber\\
\end{eqnarray}
 At every column index $j$, the explicit relation from Eq.~\eqref{Eq:Laugh_beta6} is:
\begin{eqnarray}
\label{Eq:Distance1to0}
 {2(\bf\tilde{P}}_{1j} +  {\bf P}_{1j} +   {\bf P}_{2j}) + {\bf P}_{3j} &=&0 \, ,
\end{eqnarray}
where $\mathbf{P}_{ij}$ is the coefficient in $\mathbf{P}$ of the row labeled by $\mu_i$ and column labeled by $\nu_j$. 
We have thus related a row indexed by a partition $\tilde{\mu_1}$ with $K_{ \tilde{\mu_1}}=1$ and $\Delta_{\tilde{\mu_1}}=1$ to rows indexed by partitions $\mu_1, \mu_2, \mu_3$ with distance from the cut reduced by $1$ and number of intact unit cells in $A$ reduced by $1$. 
These partitions label rows in the OEM in this example. 
A similar procedure, using the additional clustering constraint at $\beta=9$, involving the particles in the orbitals of angular momenta $2$ and $7$, can be used to relate the row of $\bf{\tilde P}$ indexed by the partition $\tilde \mu_0$ to rows in the OEM.

\subsubsection{At $k=2$:}
Let us now consider the Moore-Read state with $N=18,  N_\phi=16$, and $ L_z^{tot}=144$ and perform an orbital cut after $l_A=7$ orbitals. 
Here, we are interested in the sub-block of $\bf\tilde P$ and $\bf C$ with $N_A=8$, $L_z^A=L^A_{z,min}+\ell=L^A_{z,min}+3$. 
The occupation number configurations of $\tilde\mu_0$ and $\tilde\nu_0$ are:
\begin{eqnarray*}
n(\tilde{\mu_0}) &=& \{2020201\,|\,00100000\}  \qquad K_{\tilde{\mu_0}}=3, \Delta_{\tilde{\mu_0}}=3 \\
n(\tilde{\nu_0}) &=&  \{0000010\,|\,01020202\}  \qquad K_{\tilde{\nu_0}}=2, \Delta_{\tilde{\nu_0}}=3,
\end{eqnarray*}
where we indicate the orbital cut by the '$|$' symbol. 
Thus, $ \Delta_{\tilde\mu_0}\geq K_{\tilde\mu_0}$ and $\Delta_{\tilde\nu_0} \geq K_{\tilde\nu_0}$, and we can relate all rows and columns of the PEM to ones in the OEM.

The occupation configurations of the $(2,2)$-admissible partitions labeling the rows of $\bf\tilde{P}$ are given by:
\begin{eqnarray}
\label{Eq:MR_Na_8_Lz_3}
n(\tilde{\mu_0}) &=& \{2 0 2 0 2 0 1\, | \, 0 0 1 0 \ldots 0\} \qquad K_{\tilde{\mu_0}}=3, \Delta_{\tilde{\mu_0}}=3 \nonumber\\
n(\tilde{\mu_1}) &=& \{2 0 2 0 2 0 0\, | \, 1 1 0 0\ldots 0\} \qquad K_{\tilde{\mu_1}}=3, \Delta_{\tilde{\mu_1}}=3 \nonumber\\
n(\tilde{\mu_2}) &=& \{2 0 2 0 1 1 1 \, | \, 0 1 0 0\ldots 0\} \qquad K_{\tilde{\mu_2}}=2, \Delta_{\tilde{\mu_2}}=2 \nonumber\\
n(\tilde{\mu_3}) &=& \{2 0 2 0 1 1 0\, | \, 2 0 0 0\ldots 0\} \qquad K_{\tilde{\mu_3}}=2, \Delta_{\tilde{\mu_3}}=2 \nonumber\\
n(\tilde{\mu_4}) &=& \{2 0 1 1 1 1 1\, | \, 1 0 0  0\ldots 0\} \qquad K_{\tilde{\mu_4}}=1, \Delta_{\tilde{\mu_4}}=1\, . \nonumber\\
\end{eqnarray}
The trailing $0's$ in every occupation configuration indicate that the orbitals with $L_z=10,\ldots, 16$ are unoccupied in the partitions labeling the rows of $\tilde P$. $\Delta_{\tilde{\mu_i}}\geq K_{\tilde{\mu_i}}$ is satisfied for all $i=0,\ldots, 4$,  as required in step~\ref{Item:Result2} in Sec.~\ref{Sec:Thegeneralmethod}.

We illustrate the use of the $3-$body clustering conditions by relating the row labelled by the partition $\tilde\mu_3$ to rows labelled by partitions $\mu_j$ with distance $K_{\mu_j}=1$ from the cut. The first unit cell is the rightmost intact unit cell in $A$ in $n(\tilde{\mu_3})$. Consider the $3$-body clustering condition at $\beta$ equal to the $z$-angular momentum of the $2$ particles in the rightmost intact unit cell and a particle to the right of the cut, i.e. at $\beta=11 = 2\times 2 +7$. It is beneficial to divide the clustering condition \eqref{eq:clustering} into two terms:
\begin{align} 
\label{Eq:MRclustering_N_8_beta11}
3\left(D_{11}^{(1)} + D_{11}^{(2)}\right)|\psi\rangle=0
\end{align}
\begin{align} 
D_{11}^{(1)} &=d_2 d_2 d_7 + 2 d_2 d_3 d_6  
+ 2 d_2 d_4 d_5 + d_3 d_3 d_5 + d_3 d_4 d_4\nonumber\\
D_{11}^{(2)}&=  d_0 d_0d_{11}+2 d_0d_1d_{10} + 2 d_0d_2d_9+\ldots, 
\end{align}
 where $D_{11}^{(2)}$ contains all terms involving angular momentum orbitals 0 and/or 1.

The clustering conditions in Eq.~\eqref{Eq:MRclustering_N_8_beta11} yield a linear relation between certain coefficients in $|\psi\rangle$, for each occupation number configuration of the remaining $N-3$ particles. 
We choose the configurations $n([7,5,4,0,0]+\nu_j)$, as:
\begin{align}
|[7,5,4,0,0]+\nu_j\rangle = d_2d_2d_7 (|\tilde\mu_3 +\nu_j\rangle),
\end{align}
 where the $|\nu_j\rangle$ label the column vectors of the PEM.  
 Note that $d_2d_2d_7$ is the only term in $D_{11}^{(1)} $ that contains the angular momentum 7 orbital; all other terms have highest angular momentum less or equal 6, and thus smaller distance to the cut. 
Equivalently we can note that as $D_{11}^1$ annihilates any configuration with an occupied orbital of $z$-angular momentum greater than $7$, the first term in Eq.~\eqref{Eq:MRclustering_N_8_beta11} relates the row labeled by $\tilde\mu_3$ only to rows labelled by partitions that are dominated by $\tilde\mu_3$:
\begin{eqnarray}
\label{Eq:MR_Na_8_Lz_3_inconstraint}
n(\mu_1) &=& \{2 0 1 1 1 1 1\, | \, 1 0 0 0 \ldots 0\} \qquad K_{\mu_1}=1, \Delta_{\mu_1}=1 \nonumber\\
n(\mu_2) &=& \{2 0 1 0 2 2 0\, | \, 1 0 0 0\ldots 0\} \qquad K_{\mu_2}=1, \Delta_{\mu_2}=1 \nonumber\\
n(\mu_3) &=& \{2 0 0 2 1 2 0 \, | \, 1 0 0 0\ldots 0\} \qquad K_{\mu_3}=1, \Delta_{\mu_3}=1 \nonumber\\
n(\mu_4) &=& \{2 0 0 1 3 1 0\, | \, 1 0 0 0\ldots 0\} \qquad K_{\mu_4}=1, \Delta_{\mu_4}=1 \, .\nonumber\\
\end{eqnarray}
All the partitions above have one less intact unit cell, and smaller distance $K_{\mu_j}=K_{\tilde\mu_3}-1$ from the cut as compared to $\tilde\mu_3$.

The second operator in the clustering condition Eq.~\eqref{Eq:MRclustering_N_8_beta11} acts on states with occupation number configurations such as:
\begin{eqnarray*}
 \{4 0 0 0 1 1 0 \, &|&\, 1 0 0 0 1 0\ldots 0\} \\
 \{3 1 0 0 1 1 0 \, &|&\, 1 0 0 1 0 0\ldots 0\} \\
 \{3 0 1 0 1 1 0\, &|& \, 1 0 1 0 0  0\ldots 0\} \\
 \{3 0 0 1 1 1 0\, &|& \, 1 1 0 0 0 0\ldots 0\}\, . \\
 &\vdots&
\end{eqnarray*} 
All the above configurations have distance from the cut larger than $K_{\tilde\mu_3}=2$, and \emph{more than 2 particles} in angular momentum orbitals $0$ and $1$. Hence, they are not dominated by the root partition $\lambda_0$, and have zero weight in the model wavefunction (the corresponding row in the PEM is identically 0). 

Thus, the clustering condition at $\beta=11$ for the configuration of the remaining particles being $n([7,5,4,0,0]+\nu_j)$, yields a linear relation between the row labeled by $\tilde\mu_3$ and the rows labeled by the partitions $\mu_1,\ldots, \mu_4$:
\begin{eqnarray}
\label{Eq:Distance1}
{\bf\tilde{P}}_{3j} + 2 {\bf P}_{1j} + 2  {\bf P}_{2j} + {\bf P}_{3j} + {\bf P}_{4j} &=&0 \, ,
\end{eqnarray} 
where $\mathbf{ P}_{ij}$ is the coefficient in $\mathbf{P}$ in the row labeled by $\mu_i$ and column labeled by $\nu_j$. 
The rows labeled by $\mu_1,\ldots, \mu_4$ can in turn be related to rows in the OEM by using the clustering conditions at $\beta=7$.

\subsection{Beyond $(k,2)$-clustering states}\label{Sec:Otherstates}

Until now, we have restricted our discussions to the bosonic $(k,2)$-clustering states $\psi_{(k,2)}(z_1,\ldots, z_N)$. In this section, we generalize our results to other states with the property of clustering -- the states obtained by multiplying $(k,2)$-clustering states with $M$ Jastrow factors and the $(2,3)$-clustering Gaffnian state. We believe that our results hold for \emph{all} highest-density states uniquely defined by clustering, like, for instance, the Haffnian state. For the non-unitary states, which are supposedly bulk gapless \cite{readnonunitary,bondersonplasma}, the map relates the counting of the OES to the number of bulk quasihole states (which is equal to the counting of the PES); however, in this case the number of bulk quasiholes is \emph{not} equal to the number of the edge modes, as the edge-bulk correspondence in the energy spectrum does not hold for non-unitary states.

For the $(k,2)$-clustering states, we identified a sub-matrix of the PEM, $\bf \tilde{P}$, with the same rank as the PEM and whose smaller dimension was the number of distinct bulk quasi-hole excitations. 
We then argued, based on the lack of other symmetries in $\bf \tilde{P}$, that its rank was equal to the smaller  dimension, and that the PES counted the number of bulk quasi-hole excitations at each angular momentum. 
To generalize this argument to other clustering states, we need to first identify the special sub-matrix $\bf \tilde{P}$. 
We can then establish the bulk-edge correspondence in their entanglement spectra by slightly modifying the method used in Section~\ref{Sec:Thegeneralmethod}.
Extending the ideas in Sec.~\ref{Sec:Theproof}  is quite straightforward -- we re-define the notion of unit cell and the intact unit cell, and identify $N_c$, the number of linearly independent clustering conditions that involve the $k$ particles of an intact unit cell and one particle to the right of the cut, for a fixed occupation configuration of the remaining $N-(k+1)$ particles. $N_c=1$ for the $(k,2)$ case. Using the $N_c$ independent linear equations, we can relate a row labelled by a partition $\mu$ with $\Delta_\mu$ intact unit cells and distance to the cut $K_\mu$ to rows labelled by partitions $\mu'$, such that $\Delta_{\mu'}=\Delta_\mu-1$ and $K_{\mu'}\leq K_\mu-N_c$. Thus, in the notation of Sec.~\ref{Sec:Theproof}, when $\Delta_{\tilde{\mu_0}}\geq K_{\tilde{\mu_0}}/N_c$ and  $\Delta_{\tilde{\nu_0}}\geq K_{\tilde{\nu_0}}/N_c$, the OES and the PES have the same counting. In the thermodynamic limit, the arguments in the last paragraph in Sec.~\ref{Sec:Theproof} show that the Li-Haldane conjecture is true for these states as well. 

\subsubsection{The $(k,2)$-clustering state multiplied by Jastrow factors}
Let us consider the model wavefunction:
\begin{equation}
\label{eq:k2Jastrow}
\psi(z_1,\ldots z_N) = \psi_{(k,2)}(z_1,\ldots z_N) \prod_{i<j}(z_i-z_j)^M\, ,
\end{equation}
where $\psi_{(k,2)}(z_1,\ldots z_N)$ is the $(k,2)$-clustering state. 
In Appendix~\ref{App:tildePMrankJastrow}, we show that $\tilde{P}$ is labelled by row and column occupation configurations that obey the generalized Pauli principle: no more than $1$ particle in $M$ consecutive orbitals and no more than $k$ particles in $Mk+2$ consecutive orbitals. The unit cell has $(Mk+2)$ orbitals and the occupation configuration of the intact unit cell is $\{1(0)^{M-1}1(0)^{M-1}\ldots 1(0)^{M-1}00\}$ with $1(0)^{M-1}$ repeated $k$ times (we could succintly write the whole pattern $\{(1(0)^{M-1})^k00\}$). The exponent is the number of times the pattern in the parenthesis is repeated. In Appendix~\ref{App:fermClus}, we show that $N_c=1$ for $M=1$. More generally, $N_c=\lfloor M/2\rfloor+1$ for $k=1$ Laughlin states, and $N_c=2\lfloor M/2\rfloor +1$ for states with $k>1$. 

\subsubsection{The Gaffnian state}
The Gaffnian state is a $(2,3)$-clustering state and is a single Jack polynomial:
\begin{eqnarray}
\label{eq:gaffnian}
\psi(z_1,\ldots z_N) &=& J_{\lambda_0}^\alpha(z_1,\ldots z_N)
\end{eqnarray}
where $\alpha=-3/2$ and $n(\lambda_0)=\{200200\ldots 2002\}$.
It is described by a non-unitary CFT, the $W_2(3,5)$ model\cite{estienne-09jpamt445209, bernevig-09jpamt245206}. 
It has been suggested that the fermionic Gaffnian state is the critical state between a strong-pairing phase and a Read-Rezayi phase\cite{gaffnian}. Despite the Gaffnian being a gapless state, we can determine the counting of the PES and establish the correspondence in counting between the orbital and particle entanglement spectrum. 
The discussion in Sec.~\ref{Sec:Ptilde} and Appendix~\ref{App:tildePMrank} applies to any Jack polynomial with $(k,r)$ clustering that is a unique zero mode of a pseudopotential Hamiltonian (besides the $(k,2)$ Jacks, only one other Jack $(2,3)$  - the Gaffnian - satisfies this constraint).  $\tilde{P}$ is therefore the sub-matrix of the PEM labelled by $(2,3)$-admissible row and column occupation number configurations for the Gaffnian state. 
The unit cell has $3$ orbitals and the occupation configuration of the intact unit cell is $\{200\}$. 
We derive the clustering conditions  in Appendix~\ref{App:Gaffnian} and show that $N_c=2$ for the Gaffnian.

\subsubsection{The Haffnian state}
 The Haffnian state\cite{thesis:green01} is a $(2,4)$-clustering state, but is not a single Jack polynomial.  We cannot rigorously identify $\mathbf{\tilde P}$ for the Haffnian state, although we expect, based on our understanding of the other model states, that $\mathbf{\tilde P}$ only contains the rows and columns labeled by partitions obeying the generalized Pauli principle discussed in Ref. [\onlinecite{genPauliHaffnian}]. We have verified this numerically. The occupation configuration of the intact unit cell is $\{2000\}$. The clustering conditions are derived along the same lines as for the Gaffnian in Appendix~\ref{App:Gaffnian}, giving $N_c=3$ for the Haffnian. Whenever $\Delta_{\tilde{\mu_0}}\geq K_{\tilde{\mu_0}}/3$ and $\Delta_{\tilde{\nu_0}}\geq K_{\tilde{\nu_0}}/3$, we numerically observe that the ranks of the PEM and the OEM are equal.
 
\section{Conclusions}

In this paper we have provided a proof that the Li and Haldane natural entanglement spectrum  in the thermodynamic limit is bounded from above by the number of modes of the CFT describing the edge physics. Barring the presence of extra accidental symmetries in the system, the bound should be saturated. We showed that a part of the levels of two  different entanglement spectra, the OES and the PES,  probing different physics - that of the edge and bulk respectively - can be mapped into one another by a series of newly introduced clustering operators. This map works, for certain quantum number sectors, both in finite size and in the thermodynamic limit and provides a mathematically sound foundation for a edge-bulk correspondence \emph{in the entanglement spectra}.  Our map works for both unitary and non-unitary states that are defined as unique highest density zero-modes of Haldane pseudopotential Hamiltonians.

\section{Acknowledgements}
We acknowledge useful discussions with FDM. Haldane, B. Estienne, R. Santachiara, R. Thomale, P Bonderson, D. Arovas, XL. Qi and A. Ludwig. AC wishes to thank Ecole Normale Superieure for generous hosting. BAB and MH want to thank both Ecole Normale Superieure and Microsoft Station Q for generous hospitality. 
MH was supported by the Alexander-von-Humboldt foundation, the Royal Swedish Academy of Science, and NSF DMR grant 0952428. 
NR was supported by the Agence Nationale de la Recherche under Grant No. ANR-JCJC-0003-01, and BAB was supported by Princeton University Startup Funds, Alfred P. Sloan Foundation, NSF CAREER DMR- 095242, and NSF China 11050110420, and MRSEC grant at Princeton University, NSF DMR-0819860.

 \appendix
\section{A simple example}

Let us consider the bosonic Laughlin wavefunction of $N=4$ particles at filling $\nu=1/2$. The number of flux quanta, $N_\phi$ is $6$ and $L_z^{ tot }=12$. The wavefunction $|\psi\rangle$ can be expanded in the unnormalized basis as:
\begin{eqnarray}
\label{Eq:Sec1:psimonom}
|\psi\rangle&\equiv&\sum_\lambda b_\lambda |\lambda\rangle\nonumber\\
&=&|6,4,2,0\rangle - 2 |6,4,1,1\rangle  - 2|5,5,2,0\rangle + 4|5,5,1,1\rangle  \nonumber \\
&+& 2|6,3,2,1\rangle -2|5,4,2,1\rangle +4|5,3,2,2\rangle + 4|4,4,2,2\rangle \nonumber\\
&-& 2|6,3,3,0\rangle +2|5,4,3,0\rangle -6|4,4,4,0\rangle - 4|5,3,3,1\rangle \nonumber\\
&-& 6|6,2,2,2\rangle +4|4,4,3,1\rangle -6|4,3,3,2\rangle + 24|3,3,3,3\rangle. \nonumber\\
\end{eqnarray}
 We construct several orbital and particle entanglement matrices, and use the clustering constraints to prove the bulk-boundary correspondence in the following sub-sections.

\subsection{The orbital cut}
\label{App:ExOEMLaughlin}
Let us cut the single-particle orbital space after $l_A=3$ orbitals. Consider the blocks of the OEM at the natural number of particles in $A$, $N_A=N_{A,nat}=2$. From the above decomposition, the minimum possible angular momentum, Eq.~\eqref{eq:lzmin} for $2$ particles in A is $L_{z,min}^A=2$. At this $N_A$ and $L_z^A$, the Hilbert spaces of $A$ and $B$ are spanned by $|\mu_1\rangle=|2,0\rangle$, $|\mu_2\rangle=|1,1\rangle$ and $|\nu_1\rangle=|6,4\rangle$, $|\nu_2\rangle=|5,5\rangle$ respectively. The block \C at $N_A=2$ and $L_z^A=2$ is then given by:

\begin{equation}
\label{Eq:C_Lz_2}
\bordermatrix{ &   |6,4\rangle & |5,5\rangle \cr
 |2,0\rangle & 1 & -2 \cr
 |1,1\rangle & -2 & 4 },
\end{equation}
\noindent
 where we have indicated the states labeling the rows  and columns.  ${\bf C}_{ij}=b_{\mu_i+\nu_j}$ ( `+' as defined in Section~\ref{Sec:Notation}) and rank($\mathbf{C}$)=1. 

 The block \C with $N_A=2, L_z^A=L_{z,min}^A+1=3$, of rank $1$, is:
\begin{equation}
\label{Eq:C_Lz_3}
\bordermatrix{
 & |6,3\rangle & |5,4\rangle \cr
 |2,1\rangle & 2 & -2 } 
 \end{equation}
The block  \C at $N_A=2, L_z^A=L_{z,min}^A+2=4$, also of rank $1$, is given by:
\begin{equation}
\label{Eq:C_Lz_4}
  \bordermatrix{
& |5,3\rangle & |4,4\rangle \cr
|2,2\rangle &  4& 4  
 }
  \end{equation}
\noindent
Fig.~\ref{fig:oem_n_4_2s_6_na_2_la_3}(a) shows the numerically generated OES for the $4$ particle Laughlin state in the sphere geometry at $1/2$ filling with $N_A=2$ and $l_A=3$. The counting of the entanglement levels in the spectrum equals the ranks of \C at each $L_z^A$. 

\subsection{The particle cut}
\label{App:ExPEMLaughlin}

Let us construct the entanglement matrices for the particle cut with $N_A=2$. At the smallest possible angular momentum $L_z^A=L_{z,min}^A=2$, the PEM and OEM are identical:
\begin{eqnarray}
\label{Eq:P_Lz_2}
\bordermatrix{
 & |6,4\rangle & |5,5\rangle \cr
|2,0\rangle & 1 & -2 \cr
|1,1\rangle & -2 & 4 } \, .
\end{eqnarray}

The Hilbert space of $A$ at $L_z^A=L_{z,min}^A+1=3$ is spanned by the occupation number states $|3,0\rangle$ and $|2,1\rangle$. $|3,0\rangle$ was not a member of the Hilbert space of $A$ for the orbital cut after $l_A=3$ orbitals (discussed in the previous section), because the orbital with index $3$ belonged to $B$. The PEM at $L_z^A=3$ is given by: 
\begin{eqnarray}
\label{eq:p3}
\bordermatrix{
 & |6,3\rangle & |5,4\rangle \cr
|3,0\rangle & -2 & 2 \cr
|2,1\rangle & 2 & -2
}.
\end{eqnarray}
We see that the OEM \eqref{Eq:C_Lz_3} for $N_A=2,L_z^A=2$ is indeed a sub-matrix of the PEM, as discussed in Section~\ref{sec:pemdef}. 

As the last example, consider $L_z^A=L_{z,min}^A+2=4$. The row and the column dimension of the PEM is larger than that of the OEM in \eqref{Eq:C_Lz_4}:
\begin{eqnarray}
\label{eq:p2}
\bordermatrix{
 & |6,2\rangle  &|5,3\rangle  & | 4,4 \rangle \cr
|4,0\rangle &  1& 2 & -6\cr
|3,1\rangle &  2& -4  &4\cr
|2,2\rangle&  -6& 4  & 4
}.
\end{eqnarray}
By inspection, we see that  the OEM \eqref{Eq:C_Lz_4} is the sub-matrix consisting only of the first row and the first two columns. The rank of the PEM at $L_z^A=4$ is equal to two and greater than that of the corresponding OEM. 

\begin{figure}[htbp]
    \includegraphics[width=0.23\textwidth]{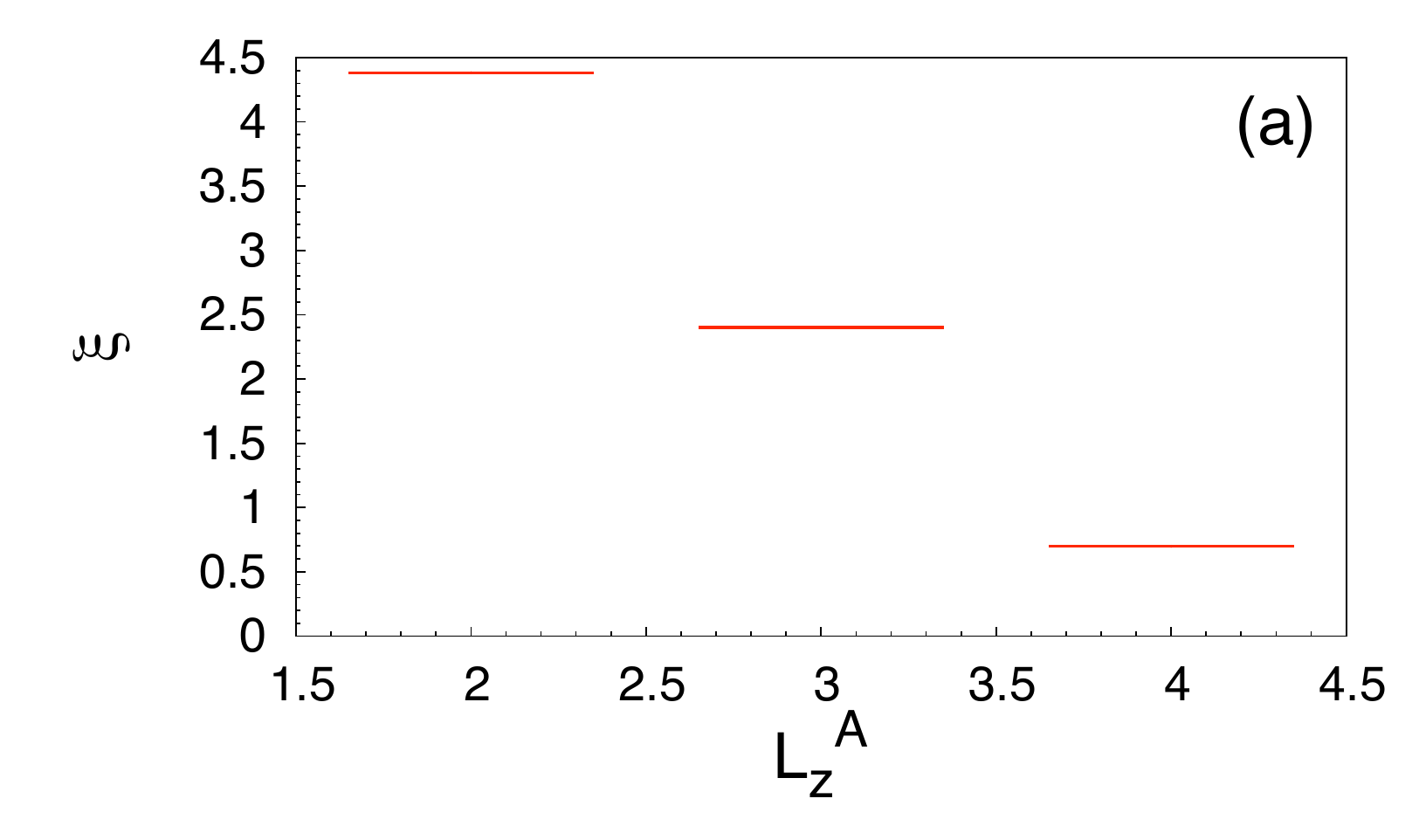}
        \includegraphics[width=0.23\textwidth]{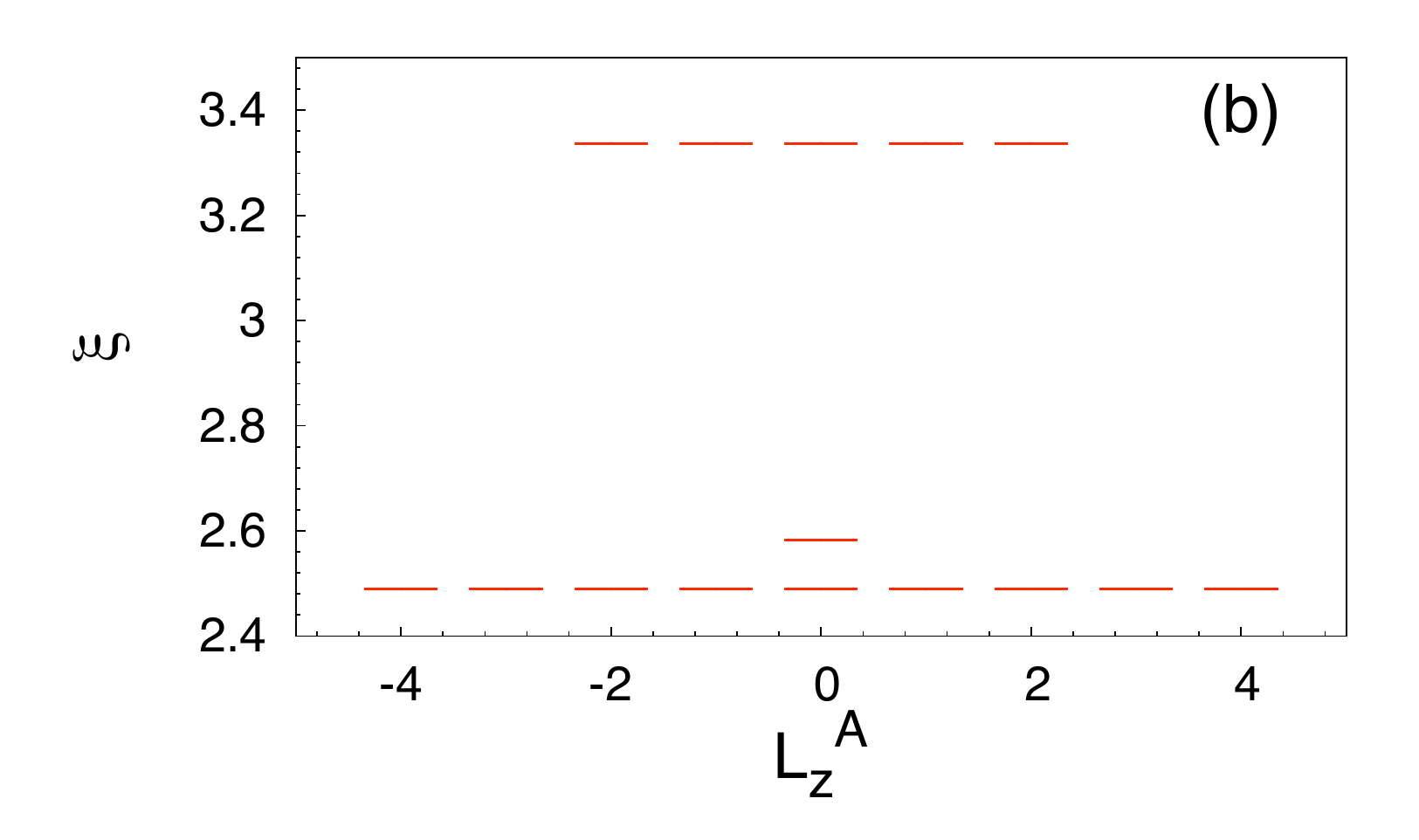}
    \caption{
 (a) Orbital entanglement spectrum of the $\nu=1/2$ Laughlin state with $N=4$, $N_A=2$, and orbital cut after $l_A=3$ orbitals. The entanglement level counting is equal to the rank of the OEM at each angular momentum. 
    (b) Particle entanglement spectrum of the $\nu=1/2$ Laughlin state with particle cut $N_A=2$. The entanglement level counting at all angular momenta $L_z^A$ is equal to the rank of the PEM.  $L_{z,min}^A$ defined in the text is $L_z^A=4$ in the plots.    }
      \label{fig:oem_n_4_2s_6_na_2_la_3}
\end{figure}

Fig.~\ref{fig:oem_n_4_2s_6_na_2_la_3}(b) shows the numerically generated PES for the $4$ particle $1/2$ Laughlin state for $N_A=2$. The counting of the spectrum agrees with the ranks calculated above.

\subsection{Relating the OES and PES counting}

 At  $L_z^A=2$, $\bf C$ (Eq.~\eqref{Eq:C_Lz_2}) and  $\bf P$ (Eq.~\eqref{Eq:P_Lz_2}) are seen to be identical. 
 There is precisely one element with a $(1,2)$-admissible occupation configuration in the Hilbert spaces of $A$ and $B$ --- $|2,0\rangle$ and $|6,4\rangle$ respectively. Thus, ${\bf\tilde{P}}=(1)$ and the three matrices $\bf P$, $\bf\tilde{P}$ and $\bf C$ are all of rank one.

At $L_z^A=3$, ${\bf\tilde{P}}=(-2)$, and is not a sub-matrix of $\bf C$ (Eq.~\eqref{Eq:C_Lz_3}). 
The matrix elements $ {\bf\tilde{P}}_{11}$ and ${\bf C}_{11}$ are the coefficients of $|6,3,3,0\rangle$ and $|6,3,2,1\rangle$ in the wavefunction $|\psi\rangle$. The 2-body clustering constraints \eqref{eq:clustering} at $\beta=3$  relate these coefficients by:
\begin{eqnarray*}
(d_3d_0 + d_2 d_1)|\psi\rangle &=& 0 \\
\Rightarrow ({\bf \tilde{P}}_{11} + {\bf C}_{11})|6,3\rangle &=& 0.
\end{eqnarray*}
This relation between the single element in $ {\bf\tilde{P}}$ and $\bf C$ proves that they have the same rank. 

The $L_z^A=4$ case is interesting. Here $\bf{\tilde{P}}$ is:
\[ \bordermatrix{
  &  |6,2\rangle&  |5,3\rangle  \cr
|4,0\rangle & 1&   2 \cr
|3,1\rangle & 2 &  -4  }. \]
$\bf{\tilde{P}}$ and $\bf C$ share the column index $|5,3\rangle$, but have no row index configurations in common. A single relation between the row vectors of $\bf {\tilde P}$ and the row labeled by the partition $[2,2]$ in $\bf C$ is provided by the $2$-body clustering conditions at $\beta=4$. 
Without another relation, we cannot relate the ranks of $\bf P$ and $\bf C$ at $L_z^A=4$. Our proof establishing the equality of ranks of the PEM and the OEM should not and is not applicable at this angular momentum, as $K_{\tilde\mu_0}=2$ and $\Delta_{\tilde\mu_0}=1$ with $\tilde\mu_0=[4,0]$.

\section{Rank of $\tilde{\bf P}$}

\subsection{$(k,2)$-clustering states}
\label{App:tildePMrank}
The matrices $\bf \tilde{P}$ and $\bf M$ were defined in Sec.~\ref{Sec:Ptilde} as the particle entanglement matrices with labels $(N_A,L_z^A)$ in the $(k,2)$-admissible occupation configuration basis and the Jack basis.  In Sec.~\ref{Sec:Ptilde}, we showed that the PEM and $\bf M$ have the same counting; in this appendix, we show that $\bf \tilde{P}$ and $\bf M$ have the same rank. This proves that the counting of the PEM equals the rank of $\bf \tilde{P}$.

Suppose we are able to show that ${\bf \tilde{P}} = {\bf D} {\bf M} {\bf D'}$, where $\bf D^T$ and $\bf D'$ are square triangular matrices with $1$'s on the diagonal and as such they  have nonzero determinant. A theorem in linear algebra states that pre/post multiplying a matrix by one of triangular form with nonzero determinant leaves its rank unchanged. Thus,  we only need to prove that ${\bf \tilde{P}} = {\bf D} {\bf M} {\bf D'}$ to conclude that rank($\bf \tilde{P}$)=rank($\bf M$).

The row and column dimensions of $\bf \tilde{P}$ and $\bf M$ are identical because every $(k,2)$-admissible partition $\mu$ labels the Jack $J_{\mu}^\alpha$. We may use partial ordering  by dominance to order the $(k,2)$-admissible row and column configurations such that if $\tilde{\mu}_k >\tilde{\mu}_i$, then $k\leq i$. 

 Consider a particular $(k,2)$-admissible partition $\tilde{\mu_i}$ ($\tilde{\nu_j}$) labeling the $i{th}$ row ($j{th}$ column) of $\bf\tilde{P}$ and $\bf M$. Let the coefficient of $|\tilde{\mu_i}\rangle$ in $|J^\alpha_{\tilde{\mu_k}}\rangle$ be ${\bf D}_{ik}$ and the coefficient of $|\tilde{\nu_j}\rangle$ in $|J^\alpha_{\tilde{\nu_l}}\rangle$ be ${\bf D'}_{lj}$. The partial ordering implies that:
\begin{eqnarray}
{\bf D}_{ik}  &=& 0 \, \textrm{ if } k>i \\
{\bf D}_{ii}  &=& 1 \\
{\bf D'}_{lj} &=& 0 \, \textrm{ if } l>j \\
{\bf D'}_{jj} &=& 1.
\end{eqnarray}
 In other normalizations of Jack polynomials, $\mathbf{D}_{ii}$ is not necessarily one, but is always non-zero. By the definition of a matrix with row-echelon form, ${\bf D}^T$ and $\bf D'$ in row-echelon form. Recall that:
\begin{equation}
\sum_{i,j} {\bf M}_{ij} |J^\alpha_{\tilde{\mu_i}}\rangle \otimes|J^\alpha_{\tilde{\nu_j}}\rangle = \sum_{i,j} {\bf P}_{ij} |\mu_i\rangle \otimes|\nu_j\rangle
\end{equation}
in every block of the full PEM. $|\mu_i\rangle$ and $|\nu_j\rangle$ are the general occupation-basis states, not just the $(k,2)$-admissible configurations. ${\bf\tilde{P}}$ is the sub-matrix of $\bf P$ labelled by $(k,2)$-admissible partitions; therefore:
\begin{eqnarray}
{\bf \tilde{P}}_{ij} &=&  \sum_{k,l} {\bf M}_{kl} \langle \tilde{\mu_i} |J^\alpha_{\tilde{\mu_k}}\rangle\langle \tilde{\nu_j} |J^\alpha_{\tilde{\nu_l}}\rangle \\
\textrm{i.e. } {\bf \tilde{P}}_{ij} &=& \sum_{k,l} {\bf D}_{ik}{\bf M}_{kl} {\bf D'}_{lj} \nonumber \\
\Rightarrow {\bf \tilde{P}} &=& {\bf D} {\bf M} {\bf D'}\, ,
\end{eqnarray}
which proves our statement that the rank of the PEM is given by the rank of the matrix of the coefficients indexed by the  $(k,2)$-admissible partitions.

\subsection{$(k,2)$-clustering states multiplied by Jastrow factors}
\label{App:tildePMrankJastrow}
We consider the PEM of states that are $(k,2)$-clustering polynomials multiplied by Jastrow factors, $\prod_{i<j}(z_i-z_j)^M$, see Eq.~\eqref{eq:k2Jastrow}.
Here we identify $\mathbf{\tilde P}$, a submatrix of the PEM with the same rank,  and find that it contains only rows (columns) that are labeled by partitions $\tilde\gamma_i$ ($\tilde \eta_j$) of $N_A$ ($N_B$) particles and angular momentum $L_z^A$ ($L_z^B$) that obey the generalized Pauli principle: there is no more than one particle in $M$ consecutive orbitals and no more than $k$ particles in $Mk+2$ consecutive orbitals. 
The total flux $N_\phi$ of the partitions $\tilde\gamma_i,\tilde\eta_j$ is equal to the total flux of the ground state $|\psi\rangle$ being cut. 

Instead of expanding $|\psi\rangle$ in terms of monomials, we can choose a different basis that incorporates all the vanishing properties of the $N_A$ ($N_B$) particles among themselves:
\begin{multline}
\langle \{z_j\}|\psi\rangle=\sum_{i,j} \mathbf{M}_{i,j}\left(J^{\alpha}_{\tilde\mu_i}\prod_{{k<k'}\atop{k,k' \in A}}(z_{k}-z_{k'})^M\right)\\ \cdot \left(J^{\alpha}_{\tilde\nu_j}\prod_{{l,l'}\atop{l,l' \in B}}(z_{l}-z_{l'})^M\right),
\end{multline}
where the Jastrow factors include only particles in part $A$ and $B$, respectively.  $\tilde\mu_i$ ($\tilde \nu_j$) are $(k,2)$-admissible partitions of $N_A$ ($N_B$) particles with angular momentum $L_z^A$ ($L_z^B$) in $2(N-1)+MN_B+1$ (for $\tilde\mu_i$) and  $2(N-1)+MN_A+1$ (for $\tilde\nu_j$) orbitals. The matrix $\mathbf{M}=(\mathbf{M}_{ij})$ has the same rank as the PEM. 

Let us, for simplicity, focus on the basis states labeling the rows of $\mathbf{M}$. The Jastrow factor can be written as a ($1,M$)-clustering Jack polynomial. Hence, both the Jack and the Jastrow state obey a dominance property. They have a root configuration with coefficient one that dominates any other configuration in the expansion in terms of occupation number states. 
This implies that also their product:
\begin{align}
J^\alpha_{\tilde\mu_i}\cdot \prod_{k<k'}(z_{k}-z_{k'})^M
\end{align}
 has a root configuration $\tilde \gamma_i$ with expansion coefficient 1, where  $(\tilde\gamma_i)_j=(\tilde\mu_i)_j+M(N-j)$. Note that the $\tilde\gamma_i$'s are precisely the configurations that label the rows of $\mathbf{\tilde P}$. 
In addition, the partitions $\tilde \gamma_i$ have the same partial ordering as the $\mu_i$, ie. if $\tilde\mu_i<\tilde\mu_j$, then $\tilde\gamma_i<\tilde\gamma_j$.
Thus, all arguments from the previous section are applicable here as well: There are row-echelon matrices $\mathbf{D}^T$ and $\mathbf{D}'$ such that $\mathbf{\tilde P}=\mathbf{DMD'}$, which proves that rank($\mathbf{\tilde P}$)=rank($\mathbf{M}$)=rank($\mathbf{P}$).

\section{The model Hamiltonian expressed as clustering operators}
\label{App:PseudoHam}
We re-write the rotationally invariant, $2$-body Haldane pseudopotential Hamiltonian, whose zero modes are $(1,2)$-clustering states, in terms of the clustering operators introduced in Sec.~\ref{Sec:Squeezingconstraints}. Recall that the single-particle orbitals in the lowest Landau level form the multiplet of $L=N_\phi/2$. In the $\hat{L_z}$ basis, any $2$-body interaction $\hat{V}$ can be expanded as:
\begin{eqnarray}
\label{Eq:Ham2bodybasis}
H=\sum_{m_i} \langle m_1,N_\phi/2; m_2, N_\phi/2 | &\hat{V}& | m_3, N_\phi/2; m_4, N_\phi/2\rangle \nonumber \\
 &&c_{m_1}^\dagger c_{m_2}^\dagger c_{m_3} c_{m_4}.
\end{eqnarray}
$c_{m_1}^\dagger$ is the creation operator of a single particle state of $L_z= m_1, L=N_\Phi/2$; $c_{m_1}^\dagger|0\rangle = |m_1, N_\phi/2\rangle$. We first change basis as follows:
\begin{eqnarray*}
&|m_1,  N_\phi/2; m_2, N_\phi/2 \rangle =  \sum_{\ell=0}^{N_\phi} |m_1+m_2, \ell \rangle  \times \nonumber \\ &\times         \langle m_1+m_2, \ell |m_1,  N_\phi/2; m_2,  N_\phi/2 \rangle        
\end{eqnarray*}
where $\langle m_1+m_2, \ell |m_1,  N_\phi/2; m_2,  N_\phi/2 \rangle$ in the RHS are Clebsch-Gordon coefficients. For brevity of notation, we drop the label $N_\phi/2$ in the superscript. To determine the components of the model pseudopotential in the new basis, we recall that $\hat{V}$ is rotationally invariant (commutes with $|\hat{L}^2|$ and $\hat{L_z}$) and penalizes only the relative angular momentum of $0$, thus:
\begin{equation*}
\langle n_1, \ell_1 | \hat{V} | n_2, \ell_{2} \rangle = \delta _{n_1, n_2} \delta_{\ell_1, \ell_2} \delta_{\ell, N_\phi}.
\end{equation*}
The hamiltonian in Eq.~\eqref{Eq:Ham2bodybasis} can therefore be written as:
\begin{eqnarray}
\label{Eq:Hamstep2}
& H=\sum_{\beta=-N_\phi}^{N_\phi}\sum_{m_1, m_3}  \nonumber \\ &  \langle \beta, N_\phi |m_1,  N_\phi/2; \beta - m_1,  N_\phi/2 \rangle^\star \times \nonumber \\ &  \times \langle \beta, N_\phi |m_3,  N_\phi/2; \beta - m_3,  N_\phi/2 \rangle \times  \nonumber \\ &\times c_{m_1}^\dagger c_{\beta-m_1}^\dagger c_{m_3} c_{\beta-m_3}.
\end{eqnarray}
The creation and annihilation operators above create and destroy particles in the normalized single-particle orbitals. Let us denote the normalization of the single-particle orbital with $L_z=m$ by $\mathcal{N}(m)$. To move to the unnormalized basis, we make the transformation:
\begin{equation}
\label{Eq:deletionintermsofcs}
d_{m}=\mathcal{N}(m) c_m.
\end{equation} 
This set of operators is identical to the `deletion' operators defined in Sec.~\ref{Sec:ClusteringDerivation}. In spinor coordinates $(u,v)$, the wavefunction of the unnormalized orbital is:
\begin{equation}
\langle u,v|d_m^\dagger |0\rangle = u^{N_\phi/2+m}v^{N_\phi/2-m}.
\end{equation}
The Clebsch-Gordan coefficients appearing in Eq.~\eqref{Eq:Hamstep2} have the form:
\begin{eqnarray}
\label{Eq:Clebsch}
 &\langle \beta, N_\phi |m_1,  N_\phi/2; \beta - m_1,  N_\phi/2 \rangle=  \nonumber  \\ &=K \mathcal{N}({m_1}) \mathcal{N}({\beta-m_1}) \sqrt{(N_\phi-\beta)! (N_\phi+\beta)!}.
\end{eqnarray}
$K$ is independent of $\beta$ and $m_1$:
\begin{equation}
K=\left( \left(\frac{4\pi}{N_\phi+1} \right)^2 \frac{\pi^{1/4}}{\sqrt{N_\phi!} 2^{N_\phi} \sqrt{(N_\phi-1/2)!} } \right)^2
\end{equation}
 Substituting Eq.~\eqref{Eq:Clebsch} and \eqref{Eq:deletionintermsofcs} in Eq.~\eqref{Eq:Hamstep2}:
\begin{multline}
\label{Eq:Hamstep3}
H=\sum_{\beta=-N_\phi}^{N_\phi}\sum_{m_1,m_3} K^2 (N_\phi-\beta)! \\ (N_\phi+\beta)!  d_{m_1}^\dagger d_{\beta-m_1}^\dagger d_{m_3} d_{\beta-m_3}.
\end{multline}
Comparing the equation above with the one in the text Eq.~\eqref{Eq:Pseudoham}, we see that $f(\beta)= (N_\phi-\beta)! (N_\phi+\beta)!$.

\section{Two examples of clustering constraints}
\label{App:ClusteringConstraintsExamples}

We write down the explicit relations imposed by the $(k+1)$-body clustering constraints discussed in Sec.~\ref{Sec:Squeezingconstraints} on the coefficients of small wavefunctions at $k=1,2$. Let us first consider an example at $k=1$, i.e. the $1/2$ Laughlin states. The clustering conditions are 2-body:
\begin{align}\label{eq:Lclustering}
\sum_{i=0}^\beta d_{\beta-i}d_i |\psi\rangle &=0\, , & \mbox{ for } \beta&=0,1,\ldots, L_z^{tot} \, .
\end{align}
Consider the $N=3$ , $L_z^{tot}=6$ wavefunction in the infinite plane geometry in which the number of orbitals is not restricted to $N_\phi +1=5$ as in the case of the sphere .  The Hilbert space is spanned by 7 partitions, $\{\lambda_i, i=\ldots 7\}$. Their corresponding coefficients in $|\psi\rangle$ are $\{b_i, i=1\ldots 7\}$:
\begin{eqnarray*}
|\psi\rangle &=& b_1 |6,0,0\rangle + b_2 |5,1,0\rangle + b_3 |4,2,0\rangle + b_4|3,3,0\rangle \\
&&  + b_5 |4,1,1\rangle + b_6 |3,2,1\rangle + b_7 |2,2,2\rangle\, .
\end{eqnarray*}
The relations at $\beta=0,1$ respectively are:
\begin{eqnarray*}
b_1 |6,0,0\rangle  = 0 &\Rightarrow & b_1=0 \\
b_2 |5,1,0\rangle  = 0 &\Rightarrow & b_2=0\, .
\end{eqnarray*}
Thus, the clustering constraints assign zero weight to $\lambda_1$ and $\lambda_2$, which are not dominated by the root partition  $\lambda_3=[4,2,0]$ ($n(\lambda_3)=\{10101\}$). 
The values of $\beta$ from 2 to 6 generate a set of 5 linearly dependent equations that fix 4 out of the 5 remaining coefficients. All the relations obtained are shown in Table~\ref{tab:laughlin3clustering}. 
The solution in terms of the coefficient of the root partition $b_3$ is $\{b_1,b_2,b_3,b_4,b_5,b_6,b_7\}=\{0,0,b_3,-2b_3,-2b_3,2b_3,-6b_3\}$.
\begin{table}[htbp]\footnotesize
\caption{ Possible occupation configurations and the clustering conditions for the $N=3, \nu=1/2$ Laughlin state at $L_z^{tot}=6$ on the infinite plane (no restriction to the number of orbitals. On the sphere the first two configurations have zero weight and the last two orbitals are missing as $N_\phi=4$}
  \begin{tabular}{cc|cc}\label{tab:laughlin3clustering}
 Coefficient of $m_\mu$ & $n(\mu)$ &  & Constraint \\
  $b_1$ & \{2000001\} & $\beta=0$: & $b_1=0$ \\
  $b_2$ & \{1100010\} & $\beta=1$: & $b_2=0$ \\
  $b_3$ & \{1010100\} & $\beta=2$: & $2b_3+b_5=0$ \\
  $b_4$ & \{1002000\} & $\beta=3$: & $b_4+b_6=0$ \\
  $b_5$ & \{0200100\} & $\beta=4$: & $2b_6+2b_3+b_7=0$ \\
  $b_6$ & \{0111000\} & $\beta=5$: & $b_5+b_6=0$ \\
  $b_7$ & \{0030000\} & $\beta=6$: & $2b_3+b_4=0$ \\
  \end{tabular}
\end{table}

 The bosonic Moore-Read state is the clustering polynomial at $k=2$. The clustering conditions involve $3$ particles:
 \begin{align}\label{eq:MRclustering}
\sum_{i,j=0}^\beta d_{\beta-i-j}d_i d_j |\psi\rangle &=0\, , & \mbox{ for } \beta&=0,1,\ldots,L_z^{tot} \, .
\end{align}

Consider the $6$-particle wavefunction with $L_z^{tot}=12$. Eq.~\eqref{eq:MRclustering} for $\beta=0$ ensures that the weight of the partitions $[4,4,4,0,0,0]$, $[5,4,3,0,0,0] \ldots [12,0,0,0,0,0]$ not dominated by $[4,4,2,2,0,0]$ is zero in the wavefunction. The number of such partitions whose coefficients are set to zero at $\beta=0$ is the number of partitions of 12 into at most 3 parts. Similarly, the constraints at $\beta=1$ set the weights of the partitions $[4,4,3,1,0,0]$, $[5,4,2,1,0,0], \ldots$ $[11,1,0,0,0,0]$ (the number of such partitions is the number of partitions of 11 into at most 3 parts) in the wavefunction to zero. The linear dependence of the set of constraints in Eq.~\eqref{eq:MRclustering} is apparent in the fact that the coefficient of the partition $[7,4,1,0,0,0]$  is set to zero by a constraint at $\beta=0$ and one at $\beta=1$. The constraints at $\beta=11,12$ are also seen to give identical relations to those at $\beta=0,1$ for this example. The configurations   $[5,4,3,0,0,0] \ldots [12,0,0,0,0,0]$  are only allowed in an infinite plane geometry. On the sphere, they would involve more orbitals than $N_\phi+1=5$ existent ones and would not appear in the Hilbert space of the decomposition of the Moore-Read ground-state. The configurations $[4,4,4,0,0,0]$ and $[4,4,3,1,0,0]$ appear on the sphere but, due to the same reason as on the infinite plane -- that they are not squeezed from the root partition --  have zero weight. 

The $16$ partitions dominated by the root partition $[4,4,2,2,0,0]$ and their corresponding coefficients in $\psi$ are shown in the second and first column of Table~\ref{tab:mr6clustering} respectively.  Let us discuss the 3-body clustering at $\beta=4$ in more detail:
\begin{equation}
\label{Eq:MR3clusteringbeta4}
3(d_4d_0d_0+2 d_3d_1d_0+d_2d_2d_0+d_2d_1d_1)|\psi\rangle=0\, .
\end{equation}
The four terms in Eq.~\eqref{Eq:MR3clusteringbeta4} individually are:
\begin{align}
\label{Eq:MR3clusteringbeta44terms}
d_4d_0d_0|\psi\rangle&= b_1|4,2,2\rangle+b_3|3,3,2\rangle\nonumber\\
 d_3d_1d_0|\psi\rangle&=b_6|4,3,1\rangle+b_7|4,2,2\rangle+b_8|3,3,2\rangle\nonumber\\
d_2d_2d_0|\psi\rangle&=b_1|4,4,0\rangle+b_7|4,3,1\rangle+b_{12}|4,2,2\rangle\nonumber\\
&+b_{11}|3,3,2\rangle\nonumber\\
d_2d_1d_1|\psi\rangle&=b_2|4,4,0\rangle+b_9|4,3,1\rangle+b_{10}|4,2,2\rangle\nonumber\\
&+b_{14}|3,3,2\rangle\, .
\end{align}
The right-hand-side of each of the four terms above is a linear combination of different occupation configurations of $3$ bosons  with total angular momentum $L_z^{tot}-\beta=8$. Since different occupation configuration states are orthogonal to each other, Eq.~\eqref{Eq:MR3clusteringbeta4} can only be satisfied if the coefficient in front of every  non-interacting many-body state is zero. Thus, we obtain four constraints on the coefficients from each of the four occupation configurations in Eq.~\eqref{Eq:MR3clusteringbeta44terms}:
\begin{align}
|4,2,2\rangle&: b_1+2b_7+b_{12}+b_{10}=0\nonumber\\
|3,3,2\rangle&: b_3+2b_8+b_{11}+b_{14}=0\nonumber\\
|4,3,1\rangle&: 2b_6+b_7+b_9=0 \nonumber\\
|4,4,0\rangle&: b_1+b_2=0.
\end{align}
The last relation also arises from the clustering constraint at $\beta=2$. 

All the relations imposed by the clustering constraints at $\beta=2,\ldots, 6$ are shown in Table~\ref{tab:mr6clustering}. Although not obvious, in this case as in the previous, the dimension of the null space of Eq.~\eqref{eq:MRclustering} is $1$. This can be analytically proved by realizing that the Moore-Read state is the densest unique ground-state of a Haldane pseudopotential Hamiltonian which can be written in terms of the clustering operators

\begin{table}[htbp]\footnotesize
\caption{Possible occupation configurations and the clustering conditions for the $N=6$ MR state at $L_z^{tot}=12$}
  \begin{tabular}{p{0.6cm}l | ll} \label{tab:mr6clustering}
  & $n(\mu)$ &  & Constraint \\
$b_{1}$ & \{20202\} &  \multirow{2}{*}{$\beta=2$:}& $ b_1+b_2=0 $ \\
$b_{2}$ & \{12102\} & & $b_3+b_6=0$\\ \cline{3-4}
$b_{3}$ & \{20121\} &  \multirow{2}{*}{$\beta=3$:}&$ 3b_3+6 b_7+b_9=0 $ \\
$b_{4}$ & \{20040\} & & $3b_4+ 6b_8+b_{13}=0$\\ 
$b_{5}$ & \{04002\} & &$6b_2+b_5=0$\\ \cline{3-4}
$b_{6}$ & \{12021\} & \multirow{2}{*}{$\beta=4$:} &$ b_1+2b_7+b_{10}+b_{12}=0 $ \\
$b_{7}$ & \{11211\} & & $b_3+2 b_8+b_{11} +b_{14}=0$\\ 
$b_{8}$ & \{11130\} & &$2 b_6+b_{7} +b_{9}=0$\\ \cline{3-4}
$b_{9} $ & \{03111\} &  \multirow{3}{*}{$\beta=5$:}&$ 2b_2+2b_7+b_9+b_{10}=0 $ \\
$b_{10}$ & \{02301\}& & $2 b_7+2b_{11}+b_{14}+ b_{15}=0$\\
$b_{11}$ & \{10320\} & & $2 b_6+2b_8+b_{13}+b_{14}=0$\\ 
$b_{12}$ & \{10401\}& & $2b_3+b_6+b_7=0$\\ \cline{3-4}
$b_{13}$ &\{03030\} & \multirow{5}{*}{$\beta=6$:}&$ 6 b_1+3b_2+6b_7+3b_3+b_{12}=0 $ \\
$b_{14}$ & \{02220\} & & $6b_2+3b_5+6b_9+3b_6+b_{10}=0$\\
$b_{15}$ &\{01410\} &  & $6b_3+3b_6+6b_8+3b_4+b_{11}=0$\\
$b_{16}$ & \{00600\}&  &$6b_7+3b_9+6b_{14}+3b_8+b_{15}=0$\\
& & &$6b_{12}+3b_{10}+6b_{15}+3b_{11}+b_{16}=0$ 
\end{tabular}
\end{table}

\section{Proof of step~\ref{Item:Result2} in Sec.~\ref{Sec:Thegeneralmethod} }
\label{App:KDeltaofkr}

We now prove the statement in Step~\ref{Item:Result2} of Section~\ref{Sec:Thegeneralmethod}--
 $K_{\tilde\mu_0}\leq \Delta_{\tilde\mu_0}$ implies that $K_{\tilde\mu}\leq\Delta_{\tilde\mu}$ for all $(k,2)$-admissible partitions $\tilde\mu$ that are dominated by $\tilde\mu_0$. We defined $\tilde\mu_0$ to be the partition that dominates all other $(k,2)$-admissible partitions at given $N_A$, $L_z^A$, (Eq.~\eqref{eq:tildemu0}):
 \begin{align}
 n(\tilde{\mu_0})&= \underbrace{k0\ldots k0}_{2\lfloor (N_A-1)/k\rfloor } x\underbrace{0\ldots01}_{\ell }0\ldots 0,
 \end{align}
where $0\leq x<k$ is fixed by the total particle number being $N_A$. We are given that $\Delta_{\tilde\mu_0}\geq K_{\tilde\mu_0}$.  The case when $K_{\tilde{\mu_0}}=0$ is trivial. All $(k,2)$-admissible partitions have distance from the cut $0$ and at least $0$ intact unit cells; therefore $K_{\tilde\mu}\leq\Delta_{\tilde\mu}$ for all $(k,2)$-admissible partitions $\tilde\mu$.

When $K_{\tilde{\mu_0}}>0$, we prove the required statement for all $(k,2)$-admissible partitions by showing that $K_{\tilde\mu}\leq\Delta_{\tilde\mu}$ for all $(k,2)$-admissible partitions $\tilde{\mu}<\tilde\mu_{0}$ at every $K_{\tilde{\mu}}$.  Let us construct the partition $\mu$ (not necessarily ($k,2$)-admissible) at the given distance $K_{\tilde{\mu}}=K_{\mu}>0$ that is dominated by all $(k,2)$-admissible partitions $\tilde{\mu}$ at the same distance. Assume that the orbital to the left of the cut is unoccupied, i.e. $n_{l_A-1}(\tilde\mu_0)=0$. If the number of particles to the right of the cut in $\tilde\mu_0$, $N_r(\tilde\mu_0)$, is equal to one then the occupation number configuration of $\mu$ is given by: 
\begin{align}
n(\mu) &= \underbrace{k0\ldots k0}_{2\Delta_{\mu}} \underbrace{(k-1)1\ldots (k-1)1}_{2(\Delta_{\tilde\mu_0}-\Delta_{\mu})}  \underbrace{x 0 \ldots 0}_{l_A-2\Delta_{\tilde\mu_0}} | \underbrace{0  \ldots 0 1}_{K_{\mu}}\, ,
\end{align}
where we denote the orbital cut  by `$|$' in the occupation configuration.
For $N_r(\tilde\mu_0)>1$, $n(\mu)$ is: 
\begin{align}
n({\mu}) &= \underbrace{k0\ldots k0}_{2\Delta_{\mu}} \underbrace{(k-1)1\ldots (k-1)1(k-1)0}_{2(\Delta_{\tilde\mu_0}-\Delta_{\mu})}   |X\ldots X,
\end{align}
where the sequence $ X\ldots X$ denotes the occupation configuration of $(N_r(\tilde\mu_0)+1)$ particles at distance $K_{\mu}$ that is maximally squeezed. 

As compared to $n(\tilde{\mu}_{0})$, the $z$-angular momentum of the particles to the left of the cut in $n({\mu})$ is increased by $\Delta_{\tilde\mu_0}-\Delta_{\mu}$, while that of the particles to the right of the cut is reduced by $K_{\tilde\mu_0}-K_{\tilde\mu}$. Since $n({\mu})$ has the same total $z$-angular momentum as $n(\tilde{\mu}_{0})$:
\begin{eqnarray*}
\Delta_{\tilde\mu_0}-\Delta_{\mu}&=&K_{\tilde\mu_0}-K_{\mu} \\
\Delta_{\tilde\mu_0}\geq K_{\tilde\mu_0} &\Rightarrow& \Delta_{\mu}\geq K_{\mu} .
\end{eqnarray*}
As every $(k,2)$-admissible partition $\tilde{\mu}$ with distance $K_{\tilde\mu}=K_{\mu}$ that dominates $\mu$ has \emph{at least} $\Delta_{\mu}$ intact unit cells:
\begin{align}
\Delta_{\tilde\mu}\geq \Delta_{\mu} ,&  \, \, \,  K_{\tilde\mu}=K_{\mu} &
\Rightarrow \Delta_{\tilde\mu}&\geq K_{\tilde\mu} 
\end{align}
at every distance from the cut.

The argument for $n_{l_A-1}(\tilde\mu_0)\neq 0$ is identical to the one described above. The only difference lies in the form of $n({\mu})$:
\begin{align}
n(\mu) &= \underbrace{k0\ldots k0}_{2\Delta_{\mu}} \underbrace{(k-1)1\ldots (k-1)1}_{2(\Delta_{\tilde\mu_0}-\Delta_{\mu})}  0 | X\ldots X,
\end{align}
where the sequence $X\ldots X$ is the maximally squeezed configuration of $x+1$ particles (for $N_r(\tilde\mu_0)=1$) respectively $k+N_r(\tilde\mu_0)$ (for $N_r(\tilde\mu_0)>1$) at distance $K_{\mu}$.

\section{Proof of step~\ref{Item:Result1} in Sec.~\ref{Sec:Thegeneralmethod} }\label{app:step1}
\subsection{Effect of dominance on the distance from the cut}\label{App:dominance_distance}
We show that dominance, {\it i.e.} $\mu> \mu'$ implies that  the distance to the cut $K_\mu\geq K_{\mu'}$,  or that squeezing cannot increase the distance from the cut.
The property of dominance is defined by:
\begin{align}
\mu>\mu' &\Rightarrow \sum_{i=1}^n \mu_i \geq \sum_{i=1}^n\mu_i'
\end{align}
for all $n\leq N$. 
Recall that $\mu_i\geq \mu_j$ for $i<j$, where $\mu_i$ and $\mu_j$ are the components of the partition $\mu$. Let us denote the number of particles to the right of the cut for any partition $\mu$ by $N_r(\mu)$. The distance from the cut $K_\mu$ can then be rewritten as:
\begin{align}
K_\mu&=\sum_{m=l_A}^{N_\phi} n_m(\mu)(m - l_A + 1) \nonumber\\
&=\sum_{i=1}^{N_r(\mu)}(\mu_i-l_A+1)\, .
\end{align}
When comparing the total distances for two partitions, $\mu$ and $\mu'$, there are three possibilities, $N_r(\mu)=N_r(\mu')$, $N_r(\mu)>N_r(\mu')$ and $N_r(\mu)<N_r(\mu')$. We will discuss them in that order: 
\begin{itemize}
\item  $N_r(\mu)=N_r(\mu')$:
\begin{align}
\mu>\mu' &\Rightarrow \sum_{i=1}^{N_r(\mu)}\mu_i\geq \sum_{i=1}^{N_r(\mu)}\mu_i'\nonumber\\
\Rightarrow &\underbrace{\sum_{i=1}^{N_r(\mu)}(\mu_i-l_A+1)}_{=K_\mu}\geq\underbrace{\sum_{i=1}^{N_r(\mu)}(\mu_i'-l_A+1)}_{=K_{\mu'}}
\end{align}
Thus, $K_\mu\geq K_{\mu'}$. 

\item $N_r(\mu)>N_r(\mu')$:
\begin{align}
\mu>\mu'&\Rightarrow \sum_{i=1}^{N_r(\mu')}\mu_i\geq\sum_{i=1}^{N_r(\mu')}\mu_i'\nonumber\\
\Rightarrow &\sum_{i=1}^{N_r(\mu')}(\mu_i-l_A+1)\geq\underbrace{\sum_{i=1}^{N_r(\mu')}(\mu_i'-l_A+1)}_{=K_{\mu'}}\, .
\end{align}
 As $\mu_i \geq l_A$ for all particles to the right of the cut, $K_\mu=\sum_{i=1}^{N_r(\mu)}(\mu_i-l_A+1)>\sum_{i=1}^{N_r(\mu')}(\mu_i-l_A+1)$. This shows that $K_{\mu}>K_{\mu'}$.

\item $N_r(\mu)<N_r(\mu')$: 
\begin{align}
\mu>\mu'&\Rightarrow \sum_{i=1}^{N_r(\mu')}\mu_i\geq \sum_{i=1}^{N_r(\mu')}\mu_i'\nonumber\\
\Rightarrow &\underbrace{\sum_{i=1}^{N_r(\mu)}(\mu_i-l_A+1)}_{=K_\mu}+\underbrace{\sum_{i=N_r(\mu)+1}^{N_r(\mu')}(\mu_i-l_A+1)}_{\leq0}\nonumber\\
&\geq\sum_{i=1}^{N_r(\mu')}(\mu_i'-l_A+1)=K_{\mu'}\, .
\end{align}
The second term must be $\leq 0$, as the particles to the left of the cut have angular momentum $\mu_i<l_A$. 
 It is strictly negative if at least one of the $\mu_i$ for $N_r(\mu)<i \leq N_r(\mu')$ is smaller that $(l_A-1)$. 
\end{itemize}
 Thus, $K_\mu\geq K_{\mu'}$ for every $\mu'$ that is dominated by $\mu$. 

\subsection{Effect of clustering conditions}\label{App:clustering}
 We show that the $(k+1)$-body clustering conditions presented in the body of the paper (Eq.~\eqref{eq:clustering}) relate the partitions $\mu$ with $\Delta_\mu>0$ intact unit cells and distance $K_\mu>0$ from the cut to partitions $\mu'$ with number of intact unit cells given by $\Delta_\mu-1$ and distance from the cut by $K_{\mu'}<K_{\mu}$. 

Let us consider an arbitrary partition $\mu$ with $\Delta_\mu$ intact unit cells  ($2 \Delta_{\mu}$ orbitals) and distance $K_\mu$:
\begin{align}
\label{eq:dompart}
n(\mu)= &\{\underbrace{k0\ldots k0}_{2\Delta_\mu} \underbrace{x\ldots x}_{l_A-2\Delta_\mu} | \underbrace{x\ldots x}_{\leq K_\mu}0\ldots 0\}\, ,
\end{align}
where we placed the orbital cut after $l_A$ orbitals. In order to keep the discussion general, we denote an arbitrary occupation number configuration by the sequence $x\ldots x$.  \footnote{The actual occupation number configuration is not generally known or important to the proof. }
For the orbitals to the right of the cut (with angular momentum $\geq l_A$) two examples of such configurations with distance from the cut, $K_\mu$,
  are:
\begin{align}\label{eq:examplex...x}
&\{\underbrace{k0\ldots k0}_{2\Delta_\mu} \underbrace{x\ldots x}_{l_A-2\Delta_\mu} |   \, \underbrace{0\ldots 0}_{K_\mu-1} 1 0\ldots 0\} \nonumber\\
&\{\underbrace{k0\ldots k0}_{2\Delta_\mu} \underbrace{x\ldots x}_{l_A-2\Delta_\mu} |  \, K_\mu \, 0\ldots 0\}\, . 
\end{align}

Let us now analyze the clustering condition that involve the $k$ particles of the rightmost intact unit cell and the rightmost particle to the right of the cut in the partition $\mu$ \eqref{eq:dompart}. 
We choose $\beta=2(\Delta_\mu-1)+\mu_1$ and require the remaining $N_A-(k+1)$ particles to occupy the same orbitals as in $n(\mu)$.  
In particular, the occupation configuration of the remaining particles has $\Delta_{\mu}-1$ intact unit cells. 
 The clustering condition relates $\mu$ only to partitions that are dominated by a partition $\mu'$ of the form \footnote{Any other configuration has zero weight in the model wavefunction, since it is not dominated by the root partition \eqref{eq:rootpartition}}:
\begin{align}
n(\mu')= &\{\underbrace{k0\ldots k0}_{2\Delta_\mu-2}(k-1)1 \underbrace{x\ldots x}_{l_A-2\Delta_\mu}  | \tilde x\ldots \tilde x0\ldots 0\}
\end{align}
where $\tilde x\ldots \tilde x$ is used to indicate an occupation number configuration where the rightmost particle to the right of the cut is moved  to the left by one orbital. 
The distance from the cut is reduced by one: $K_{\mu'}=K_{\mu}-1$. 
For our examples in Eq. \eqref{eq:examplex...x}, the dominating partition is given by:
\begin{align}
n(\mu')= &\{\underbrace{k0\ldots k0}_{2\Delta_\mu-2}(k-1)1 \underbrace{x\ldots x}_{l_A-2\Delta_\mu}  |  \, \underbrace{0\ldots 0}_{K_\mu-2} 1 0\ldots 0\}
\end{align}
for the configuration of the first line of Eq. \eqref{eq:examplex...x}, and:
\begin{align}
n(\mu')= &\{\underbrace{k0\ldots k0}_{2\Delta_\mu-2}(k-1)1 \underbrace{x\ldots x}_{l_A-2\Delta_\mu}  |  \, (K_\mu-1) \, 0\ldots 0\}\,
\end{align}
for the configuration in the second line of Eq. \eqref{eq:examplex...x}. 

Using the results from Appendix~\ref{App:dominance_distance}, we conclude that all partitions $\mu'\neq \mu$ involved in the clustering condition have $\Delta_{\mu'}=\Delta_\mu-1$ intact unit cells and distance from the cut $K_{\mu'}\leq K_{\mu}-1$.  
The ($k+1$)-body clustering condition yields one constraint on the rows corresponding to all the involved partitions. Thus we have shown that the row labeled by $\mu $ can be written as a linear combination of the rows labeled by partitions $\mu'$ with $K_{\mu'}<K_{\mu}$ and one less intact unit cell.

\subsection{Relating PEM rows to OEM rows}\label{App:pem->oem}
Let us assemble the results of the previous appendices to prove the following statement: any PEM row corresponding to a partition $\mu$ with $K_\mu\leq \Delta_\mu$ is linearly dependent on rows belonging solely to the OEM. 
We prove this statement by induction, starting with a row partition $\mu$ with $K_\mu=1$ and $\Delta_\mu\geq 1$. Such a row partition is necessarily of the form:
\begin{align}
\label{app:pemrow}
\{\underbrace{k0\ldots k0}_{2\Delta_\mu} \underbrace{x\ldots x}_{l_A-2\Delta_\mu} | 1 0\ldots 0\}\,  .
\end{align}
Using the $(k,2)$ clustering constraint for $\beta=2k(\Delta_\mu-1)+l_A$ and fixing the occupation configuration of the remaining $N-(k+1)$ particles to be:
\begin{align}
\{ \underbrace{k0\ldots k0}_{2\Delta_{\mu}-2}00\underbrace{x\ldots x}_{l_A-2\Delta_\mu} |0\ldots 0\}
\end{align} 
 the row partition \eqref{app:pemrow} can be related to row partitions of the OEM only. Thus, this row can be written as a linear combination of the rows of the OEM, independent on $\Delta_\mu$ as long as $\Delta_\mu\geq 1$.

For the induction hypothesis, let us now assume that all row partitions $\lambda_j$ at fixed $K_{\lambda_j}\equiv K_\lambda>1$ with  $\Delta_{\lambda_j}\geq K_{\lambda}$ can be written as linear combinations of the rows of the OEM. 
Now consider a row partition $\mu$ with $K_\mu=K_{\lambda}+1$ and $\Delta_\mu\geq K_\mu$. In Appendix~\ref{App:clustering} we have showed  that a clustering condition involving any of the particles to the right of the cut and the $k$ particles of the rightmost intact unit cell (to the left of the OEM cut) relates this partition to the partitions $\mu'$ with $K_{\mu'}<K_\mu$ and $\Delta_{\mu'}=\Delta_\mu-1$. 
This implies that the  row partition $\mu$ is a linear combination of the row partitions $\lambda_j$. Using the induction hypothesis, the row partitions $\lambda_j$ can be written as linear combinations of the rows of the OEM. 
Therefore, the row partition $\mu$ can also be written as a linear combination of the OEM, which proves the conjecture.

\section{More clustering conditions}
 We derive the clustering conditions of two particular states that are uniquely defined by vanishing properties distinct from $(k,2)$. It should be possible to extend the general ideas here to other model states. For $r>3$, or for $r=3$ and $k>2$, the clustering conditions derived by requiring that the polynomial wavefunction dies with the $r$'th power of the difference between the coordinates of $k+1$ particles do not uniquely define the wavefunction\cite{Simon:2007kx}. 

\subsection{ Gaffnian state}\label{App:Gaffnian}
The bosonic Gaffnian state is a $(2,3)$-clustering state\cite{bernevig08prl246806,gaffnian}. It vanishes as the third power between the coordinate of a cluster of two particles and that of a third particle approaching the cluster:
\begin{align}
\psi_{(2,3)}(z_1,z_1,z_3,\ldots, z_N)&\propto (z_{1,3})^3, \, \,  \mbox{ for } z_{1,3}\rightarrow 0\, ,
\end{align}
where we define $z_{i,j}=z_i-z_j$. Therefore: 
\begin{align}\label{eq:Gaffnianconstraint}
\lim_{z_{1,2},z_{1,3}\rightarrow 0} (z_{1,3})^{-\alpha} \psi_{(2,3)}(z_1,z_2,z_3,\ldots, z_N)&=0, 
\end{align}
for $\alpha=0,1$ and 2. 
 Exactly as we did in Sec.~\ref{Sec:Squeezingconstraints}, we separate the coordinates $z_1,z_2 $ and $z_3$ from the rest and rewrite the Gaffnian wavefunction as:
\begin{multline}\label{eq:Gaffnianexpanded}
\psi_{(2,3)}(z_1,\ldots, z_N) \\
=\sum_{l_1,\ldots, l_3} \left(\prod_{j=1}^3 z_j^{l_j}\right) \langle z_4,\ldots, z_N | \prod_{j=1}^3 d_{l_j}|\psi \rangle
\end{multline}
where the  $d_{l_j}$'s are the destruction operators defined in Section~\ref{Sec:Squeezingconstraints}. Expanding $z_3^{l_3}$ as:
\begin{align}
z_3^{l_3}&=(z_1-(z_{1,3}))^{l_3}\nonumber\\
&=\sum_{j=0}^{l_3} \binom{l_3}{j} z_1^{l_3-j}(-z_{1,3})^{j}
\end{align}
and inserting \eqref{eq:Gaffnianexpanded} into Eq. \eqref{eq:Gaffnianconstraint}, we obtain the clustering constraints:
\begin{align}
\sum_{l_2,l_3} \binom{l_3}{\alpha} d_{\beta-l_2-l_3} d_{l_2}d_{l_3} |\psi\rangle =0, \, \, \forall \beta\geq \alpha
\end{align}
for $\alpha=0,1,2$. The clustering conditions at $\alpha=0$ are identical to the ones we derived for the Moore-Read state in Sec.~\ref{Sec:Squeezingconstraints}, as a $(2,3)$-clustering state also satisfies $(2,2)$-clustering. The set of clustering conditions at each value of $\beta$ are linearly dependent; in fact, for each $\beta>2$, the number of linearly independent clustering relations is $N_c=2$.

\subsection{ Fermionic $(k,2)$-clustering states}\label{App:fermClus}
The fermionic counterpart of the $(k,2)$-clustered bosonic state is:
\begin{equation}\label{eq:Lconstraint}
\psi(z_1,\ldots,z_N)=\psi_{(k,2)}(z_1,\ldots,z_N) \cdot \prod_{i<j}(z_i-z_j)\, .
\end{equation}

 Let us start with the simplest example, the Laughlin state for $k=1$. 
From the form of the wavefunction, $\psi=\prod_{i<j}(z_i-z_j)^3$, we see that:
\begin{align}
\lim_{z_{1,2}\rightarrow 0} z_{1,2}^{-\alpha} \, \psi(z_1,\ldots,z_N) =0, \qquad \mbox{for } \alpha=0,1,2
\end{align}
with $z_{i,j}=z_i-z_j$. Let us introduce a fermionic deletion operator $d_i$ that destroys a fermion in angular momentum orbital $i$, analogous to the bosonic case Eq.~\eqref{eq:delop}. 
We can rewrite the wavefunction as:
\begin{align}
\psi(z_1,\ldots,z_N) &= \sum_{l_1,l_2} z_1^{l_1} z_2^{l_2} \langle z_3,\ldots, z_N| d_{l_1} d_{l_2} |\psi\rangle 
\end{align}
and expand $z_2^{l_2}$ as:
\begin{align}\label{eq:expandz_2}
z_2^{l_2} &= (z_1-z_{1,2})^{l_2} \nonumber\\
&=\sum_{j=0}^{l_2} \binom{l_2}{j} z_1^{l_2-j} (-z_{1,2})^j\, .
\end{align}
Inserting this expression of $\psi$ into Eq.~\eqref{eq:Lconstraint} and taking the limit $z_{1,2}\rightarrow 0$, the only non-vanishing contribution is for $j=\alpha(=0,1,2)$--- all others vanish trivially--- and we arrive at the clustering conditions: 
\begin{align}
0&=\sum_{l_1,l_2} \binom{l_2}{\alpha} d_{l_1}d_{l_2} |\psi\rangle
\end{align}
The condition at $\alpha=0$ is identically zero due to the anti-commutation relations of the fermionic operators. 

For $\alpha =1$, we find --- using $\beta=l_1+l_2$:
\begin{align}\label{eq:Lclust}
0&=\sum_{l=0}^\beta l d_{\beta-l}d_{l} |\psi\rangle, \qquad \mbox{for } \beta\geq 1 .
\end{align}
When applying the above conditions, one must  account for the anti-commutation of the fermionic deletion operators $d_l$.
Choosing $\alpha=2$ yields clustering constraints that are identical to those at  $\alpha=1$, up to an overall multiplicative constant. 
Thus, for the fermionic model state at $\nu=1/3$ we find only one clustering condition, Eq.~\eqref{eq:Lclust}, as in  the bosonic case. 

For $k>1$ a very similar picture emerges. The two-body clustering conditions that originates from requiring:
\begin{align}
\lim_{z_{1,2}\rightarrow 0 } \psi(z_1,\ldots, z_N) \equiv 0
\end{align}
is equivalent to Pauli exclusion statistics. 
In order to find the relevant  $(k+1)$-particle clustering condition, we  divide the wavefunction by a full Jastrow factor of the particles $z_1,\ldots, z_{k+1}$:
\begin{align}\label{eq:k+1clust}
0&\equiv \lim_{z_{1,2},\ldots, z_{1,k+1}\rightarrow 0}\left( \prod_{i<j}^{k+1} z_{i,j}^{-1}\right) \, \psi(z_1,\ldots, z_N)\, .
\end{align}
Following the same steps as in the previous subsection we find the clustering conditions:
\begin{align}\label{eq:k+1clustcond}
0&=\sum_{l_1,\ldots, l_{k+1}} \prod_{j=1}^{k+1} \binom{l_j}{j} d_{l_j} |\psi\rangle 
\end{align}
with $l_1+ l_2+ \ldots l_{k+1} =\beta $.

In principle, one can also analyze variants of Eq.~\eqref{eq:k+1clust}, where not a full Jastrow factor is divided out, and derive clustering conditions from them. 
However, the resulting conditions are identical zero due to the anti-commuting operators. 
The only non-trivial relation is the one given in Eq.~\eqref{eq:k+1clustcond}. 

In general, when multiplying the $(k,2)$-clustering model state with $M$ Jastrow factors ($M>1$), 
 we find $\lfloor M/2\rfloor$ 2-body clustering constraints, and (for $k>1$) $\lfloor M/2\rfloor $ 3-body clustering constraints, in addition to the original $(k+1)$-body clustering constraint from the model state. Thus, the total number, $N_c$, of clustering constraints is $N_c=2\lfloor M/2\rfloor+1$.

\bibliography{pemtooem}

\end{document}